\documentstyle[eqsecnum,aps,prb]{revtex}

\begin{document}
\draft
\title{The Meissner effect and the vortex structure in stacked junctions and
layered superconductors: Exact analytical results}
\author{Sergey V. Kuplevakhsky}
\address{Department of Physics, Kharkov National University,\\
61077 Kharkov, Ukraine}
\date{\today}
\maketitle

\begin{abstract}
We present an exact mathematical description of the Meissner effect and of
the vortex state in periodic thin-layer superconductor/insulator structures
with an arbitrary number of identical junctions $N-1$ ($2\leq N<\infty $,
where $N$ is the number of superconducting layers) in the presence of a
static parallel external field $H$. Based on an analytical analysis of the
coupled static sine-Gordon (SG) equations for the phase differences, we
obtain a complete classification of all possible types of physical
solutions. We prove that at $H>0$ these equations admit only Meissner
solutions and topological ''vortex-plane'' solutions. Both the types of
solutions are characterized, in general, by $\left[ \frac N2\right] $
Josephson lengths ($\left[ \frac N2\right] $ is the integer part of $\frac N2
$). We derive an explicit analytical expression for the superheating field
of the Meissner state, $H_s$, as a function of $N$ and show that $H_s$
simultaneously determines the penetration field for the vortex planes. For $%
H\ll H_s$, we obtain a closed-form analytical expression for the Meissner
field and investigate a transition to the infinite ($N=\infty $)
layered-superconductor limit. Thermodynamically stable ''vortex-plane''
solutions represent coherent chains of Josephson vortices (one vortex per
each insulating layer in a chain). Being a natural generalization of
ordinary Josephson vortices in a single junction, the vortex-plane solutions
inherit such properties of the former as periodicity along the layers and
the overlapping of states with different topological numbers. We obtain
exact analytical expressions for the self-energy of a vortex plane and for
the lower critical field $H_{c1}$. In contrast to a prevailing view, the
coupled SG equations do not possess any single-vortex solutions for $H>0$,
as well as such vortex solutions as the ''triangular lattice'' and the ''row
mode''. Thus, single-vortex solutions appear only in the limit $H=0$, for
which case we provide a detailed analytical description. The general
consideration is illustrated by two exactly-solvable examples ($N=2,3$).
Experimental implications of the results are discussed.
\end{abstract}


\section{Introduction}

We present a rigorous mathematical examination of the problem of the
Meissner effect and of the vortex structure in thin-layer Josephson-junction
stacks and layered superconductors, with a static, homogeneous external
magnetic field $H>0$ applied parallel to the layers (along the $z$ axis, see
Fig. 1.) We consider periodic systems composed of an arbitrary number $N-1$
of identical superconductor/insulator (S/I) junctions ($2\leq N<\infty $,
where $N$ is the number of S-layers, with $x$ being the layering axis),
occupying an arbitrary region $\left[ -L,L\right] $ along the $y$ axis.

We begin with the investigation of analytical properties of a finite set of
coupled static sine-Gordon (SG) equations for phase differences $\phi _n$($%
1\leq n\leq N-1$), obtained by the minimization of a microscopic Gibbs
free-energy functional.\cite{K99,K00} Equations of this type were first
derived within the framework of a phenomenological approach.\cite{SBP93}
Under corresponding redefinition of the parameters, they also apply to the
phenomenological Lawrence-Doniach (LD) model\cite{LD} with $N<\infty $.
Although particular numerical solutions to these equations were obtained in
several publications,\cite{SBP93,KW97,Kr00,Kr01} any analytical analysis of
their general properties has not been undertaken up until now.

Making use of standard methods of the theory of differential equations,\cite
{T61} we arrive at a complete classification of the solutions to the SG
equations subject to appropriate physical boundary conditions at $y=\pm L$.
In particular, we prove that for $H>0$ these equations admit only Meissner
solutions and topological ''vortex-plane'' solutions.\cite{K99,K00} Both the
types of solutions are characterized, in general, by $\left[ \frac N2\right] 
$ Josephson lengths $\lambda _{Jk}$ ($\left[ \frac N2\right] $ is the
integer part of $\frac N2$, $k=0,1,\ldots ,\left[ \frac N2\right] -1$). The
Meissner solutions persist up to a certain superheating field of the
Meissner state, $H_{sL}$. We show that the field $H_{sL}$ simultaneously
determines the penetration field for the vortex planes. For $L\gg \lambda
_{J\max }\equiv \lambda _{J0}$, we derive an explicit analytical expression
for the superheating (penetration) field $H_s\equiv H_{s\infty }$ as a
function of $N\geq 2$. For $H\ll H_s$, we obtain a closed-form analytical
expression for the Meissner local field $H_n(y)$, valid for any $N\geq 2$,
and investigate a transition to the infinite ($N=\infty $)
layered-superconductor limit.

Thermodynamically stable vortex-plane solutions, physically, represent
coherent chains of Josephson vortices (one vortex per each I-layer in a
chain), positioned in planes parallel to the coordinate axes $x$, $z$. (See
Fig. 2.) Such solutions were thoroughly studied for infinite layered
superconductors in Refs. \cite{K99,K00} . For $H_{c1}\ll H\ll H_{c2}$, they
correspond to the ''transparent state'' considered by Theodorakis within the
framework of the infinite LD model.\cite{Th90} In the case of a
double-junction stack ($N=3$), they coincide with the well-known ''in-phase
mode''.\cite{SBP93,SAUK96,GGU96,KW97} Significantly, coherent Josephson
vortex configurations that can be identified with the vortex-plane solutions
have been directly observed in experiments on artificial double-junction
stacks\cite{N96} and weakly-coupled multilayers.\cite{Yu98} Besides giving a
proof of the existence and stability of the vortex-plane solutions for $%
3\leq $$N<\infty $, we discuss their main physical properties, such as
periodicity in the $y$ direction and the overlapping of states with
different numbers of vortex planes $N_v$. In the limit $L\gg \lambda _{J\max
}\equiv \lambda _{J0}$, we derive exact analytical expressions for the
self-energy $E_v$ of the state with $N_v=1$ and for the lower critical field 
$H_{c1}$. We also investigate a physical and mathematical relationship
between vortex-plane solutions for $N\geq 3$ and the ordinary Josephson
vortices in a single junction ($N=2$).

In contrast to a prevailing view, the coupled SG equations do not possess
any single-vortex solutions for $H>0$ and $L<\infty $, as well as such
vortex solutions as the ''triangular lattice''\cite{V89,Ca91,BC91} and the
''row mode''.\cite{Kr00} Thus, single-vortex solutions appear only in the
limit $H=0$, $L=\infty $, which, physically, signifies their absolute
thermodynamic instability. In this regard, it should be emphasized that
single-vortex configurations were originally proposed for infinite layered
superconductors without an appropriate analysis of the actual conditions of
their existence.\cite{B73,CC90,Ca91,Ko93} (See Refs. \cite{F98,K99,K00} for
the criticism.) An excessive preoccupation with single-vortex configurations
in previous publications can be explained by the confusion of the problem of
finding the lowest-energy topological solutions at $H=0$ with the problem of
the minimization of the Gibbs free energy at $H>0$: Although at $H=0$, $%
L=\infty $ the energy of single-vortex configurations is lower than that of
the vortex planes, the former ones are irrelevant to the problem of
minimization because of their absolute instability at $H>0$, $L<\infty $.
For the same reason, single-vortex solutions for $H=0$, $L=\infty $ cannot
be used for estimates of the lower critical field $H_{c1}$. The absence of
single-vortex configurations at $H>0$ was established for infinite layered
superconductors by the use of exact variational methods in Refs. \cite
{K99,K00} . In order to clarify this issue for $N<\infty $, we analyze those
features of single-vortex solutions at $H=0$, $L=\infty $ which preclude
their existence at $H>0$, $L<\infty $.

Section II of the paper is devoted to exact mathematical formulation of the
problem. In section III, we derive all major mathematical and physical
results sketched above. Mathematical {\bf Lemmas 1}, {\bf 2} and the {\bf %
Theorem} of subsection IIIA form the basis for a subsequent physical
analysis in subsections IIIB-IIID. The general consideration of section III
is illustrated in section IV by two exactly-solvable examples, namely of a
single thin-layer junction ($N=2$) and of a double-junction stack ($N=3$).
(In section IV, we present a new exact solution to the single SG equation,
valid for arbitrary $H\geq 0$ and $L>0$.) The obtained results are discussed
in section V. Appendices A-C contain mathematical derivations and proofs
omitted in the main text.

\section{Formulation of the problem}

We consider a periodic structure consisting of alternating $N$
superconducting(S) and $N-1$ insulating (I) layers ($2\leq N<\infty $) in a
parallel, static, homogeneous external magnetic field $0<H\ll H_{c2}$, where 
$H_{c2}$ is the upper critical field. (See Fig. 1.) The S-layer thickness is 
$a$, and $p$ is the period; the length of the structure in the $y$ direction
is $W=2L$, and $W_z$ is the length of the structure in the $z$ direction ($%
W_z\rightarrow \infty $). We set $\hbar =c=1$ and assume that 
\begin{equation}  \label{1.9}
\frac{T_{c0}-T}{T_{c0}}\ll 1,
\end{equation}
\begin{equation}  \label{1.10}
\xi _0\ll a,
\end{equation}
\begin{equation}  \label{1.11}
a\ll \min \left\{ \zeta (T),\lambda (T),\alpha ^{-1}\xi _0\right\} ,\qquad
\alpha \equiv \frac{3\pi ^2}{7\zeta (3)}\int\limits_0^1dttD(t)\ll 1,
\end{equation}
\begin{equation}  \label{1.12}
a\ll p.
\end{equation}
Here $T_{c0}$ is the critical temperature of an isolated S-layer, $\xi _0$
is the BCS coherence length; $\zeta (T)$ and $\lambda (T)$ are,
respectively, the Ginzburg-Landau (GL) coherence length and the penetration
depth; $D(\cos \theta )$ is the incidence-angle-dependent tunneling
probability of the I-layer between two successive S-layers.

Conditions (\ref{1.9}) and (\ref{1.10}) ensure the applicability of the
GL-type expansion within each S-layer. Condition (\ref{1.11}) corresponds to
the thin S-layer limit, whereas condition (\ref{1.12}) is employed here for
the sake of mathematical simplicity only. To simplify further the
mathematical description, we introduce dimensionless units by 
\[
\frac xp\rightarrow x, 
\]
\[
\frac y{\bar \lambda _J}\rightarrow y, 
\]
\[
\frac H{\bar H_s}\rightarrow H, 
\]
where the quantities on the left-hand side are dimensional, with $\bar %
\lambda _J=\left( 8\pi ej_0p\right) ^{-1/2}$ being the Josephson penetration
depth ($j_0$ is the density of the Josephson current in a single junction
with thick electrodes) and $\bar H_s=\left( ep\bar \lambda _J\right) ^{-1}$
being the superheating (penetration) field of the infinite layered
superconductor.\cite{K99,K00} In our dimensionless units, for example, the
flux quantum is $\Phi _0=\pi $, and the lower critical field of the infinite
layered superconductor\cite{K99,K00} is $\bar H_{c1}=\frac 2\pi $.

Under the above conditions, the structure is completely described by a
closed set of self-consistent microscopic equations for the reduced modulus
of the superconducting order parameter in the $n$th S-layer $f_n(y)$ ($0\leq
f_n\leq 1$, $n=0,1,\ldots ,N-1$) and the phase difference between the $n$th
and $(n-1)$th S-layers $\phi _n(y)$ ($\phi _n\equiv \varphi _n-\varphi
_{n-1} $, $n=1,2,\ldots ,N-1$):\cite{r1} 
\[
f_0(y)-f_0^3(y)=r(T)\left[ \frac{\epsilon ^2}2\frac{d^2f_0(y)}{dy^2}+\frac{2%
\left[ H-H_1(y)\right] ^2}{\epsilon ^2f_0^3(y)}+\frac 12\left[
f_0(y)-f_1(y)\cos \phi _1(y)\right] \right] , 
\]
\[
f_n(y)-f_n^3(y)=r(T)\left[ \frac{\epsilon ^2}2\frac{d^2f_n(y)}{dy^2}+\frac{2%
\left[ H_n(y)-H_{n+1}(y)\right] ^2}{\epsilon ^2f_n^3(y)}\right. 
\]
\begin{equation}  \label{1.16}
\left. +\frac 12\left[ 2f_n(y)-f_{n+1}(y)\cos \phi _{n+1}(y)-f_{n-1}(y)\cos
\phi _n(y)\right] \right] ,\quad 1\leq n\leq N-2,
\end{equation}
\[
f_{N-1}(y)-f_{N-1}^3(y)=r(T)\left[ \frac{\epsilon ^2}2\frac{d^2f_{N-1}(y)}{%
dy^2}+\frac{2\left[ H-H_{N-1}(y)\right] ^2}{\epsilon ^2f_{N-1}^3(y)}\right. 
\]
\[
\left. +\frac 12\left[ f_{N-1}(y)-f_{N-2}(y)\cos \phi _{N-1}(y)\right] %
\right] ; 
\]
\begin{equation}  \label{1.17}
\frac{df_n}{dy}\left( \pm L\right) =0,\quad 0\leq n\leq N-1;
\end{equation}
\[
\frac 1{f_n^2(y)}\left[ H_{n+1}(y)-H_n(y)\right] -\frac 1{f_{n-1}^2(y)}\left[
H_n(y)-H_{n-1}(y)\right] -\epsilon ^2H_n(y) 
\]
\begin{equation}  \label{1.18}
=-\frac{\epsilon ^2}2\frac{d\phi _n(y)}{dy},\quad 1\leq n\leq N-1,
\end{equation}
\begin{equation}  \label{1.19}
H_0(y)=H_N(y)=H,
\end{equation}
\begin{equation}  \label{1.20}
\frac{d\phi _n}{dy}\left( \pm L\right) =2H,\quad 1\leq n\leq N-1,
\end{equation}
where 
\[
r(T)\equiv \frac{\zeta ^2(T)\alpha }{a\xi _0},\qquad \epsilon \equiv \frac{%
\sqrt{ap}}\lambda , 
\]
and the local magnetic field in the $n$th I-layer ($n-1<x<n$) is given by
\[
H_n(y)=\frac 12\int\limits_{-L}^yduf_n(u)f_{n-1}(u)\sin \phi _n(u)+H 
\]
\begin{equation}  \label{1.22}
=\frac 12\int\limits_L^yduf_n(u)f_{n-1}(u)\sin \phi _n(u)+H.
\end{equation}
Note that although in most physical situations $\epsilon \ll 1$, we will not
need the smallness of $\epsilon $ in our mathematical consideration. Because
of the obvious property $f_n(y)=f_n(-y)$, equation (\ref{1.22}) implies that
the phase differences $\phi _n$ meet the condition 
\begin{equation}  \label{1.23}
\phi _n(y)=-\phi _n(-y)+0\text{mod}2\pi .
\end{equation}

The dimensionless Gibbs free energy $\Omega (H)$, normalized via the relation
\[
\frac{4\pi \Omega (H)}{H_c^2(T)a\lambda _{J\infty }W_z}\rightarrow \Omega
(H), 
\]
in terms of the mean-field quantities $f_n$, $\phi _n$ and $H_n(y)$ has the
form
\[
\Omega (H)=\sum_{n=0}^{N-1}\int\limits_{-L}^Ldy\left[ -f_n^2(y)+\frac 12%
f_n^4(y)+\frac{r(T)\epsilon ^2}2\left( \frac{df_n(y)}{dy}\right) ^2\right. 
\]
\[
\left. +\frac{2r(T)}{\epsilon ^2f_n^2(y)}\left[ H_{n+1}(y)-H_n(y)\right]
^2+2r(T)\left[ H_n(y)-H\right] ^2\right] 
\]
\begin{equation}  \label{1.24}
+\frac{r(T)}2\sum_{n=1}^{N-1}\int\limits_{-L}^Ldy\left[
f_{n-1}^2(y)+f_n^2(y)-2f_n(y)f_{n-1}(y)\cos \phi _n(y)\right] .
\end{equation}
Here, the sum of the three terms in the first line on the right-hand side is
the condensation energy, the sum of the two terms in the second line is the
electromagnetic energy $E_{em}$, and the last term is the Josephson energy $%
E_J$. The intralayer current in the $n$th S-layer $J_n(y)$ (normalized to $%
\bar H_s$) and the density of the Josephson current between the $n$th and
the $(n-1)$th S-layers $j_{n,n-1}(y)$ (normalized to $j_0$) are given by 
\begin{equation}  \label{1.25}
J_n(y)=\frac 1{4\pi }\left[ H_n(y)-H_{n+1}(y)\right] ,\quad 0\leq n\leq N-1,
\end{equation}
and 
\begin{equation}  \label{1.26}
j_{n,n-1}(y)=2\frac{dH_n(y)}{dy}=f_n(y)f_{n-1}(y)\sin \phi _n(y),\quad 1\leq
n\leq N-1,
\end{equation}
respectively.

Assuming that the temperature range satisfies the condition of the
weak-coupling limit 
\begin{equation}  \label{1.28}
r(T)\ll 1,
\end{equation}
one can obtain a perturbative solution for $f_n$ and $\phi _n$ up to any
desired order in $r(T)$, starting from the zero-order solution to (\ref{1.16}%
), (\ref{1.17}), 
\begin{equation}  \label{1.29}
f_n=1,
\end{equation}
and the zero-order equations for $\phi _n$, 
\begin{equation}  \label{1.30}
H_{n+1}(y)-\left( 2+\epsilon ^2\right) H_n(y)+H_{n-1}(y)=-\frac{\epsilon ^2}2%
\frac{d\phi _n(y)}{dy},\quad 1\leq n\leq N-1,
\end{equation}
where $H_n(y)$ are given by (\ref{1.22}) with $f_n=1$ and satisfy the
boundary conditions (\ref{1.19}). For most applications, it is sufficient to
consider expressions for physical quantities only in leading order in $r(T)$%
. Thus, for example, substituting (\ref{1.29}) and the solution of (\ref
{1.30}) into (\ref{1.24}) immediately yields a first-order expansion for the
Gibbs free energy, because first-order corrections to the
condensation-energy term cancel out.

A detailed mathematical analysis of Eqs. (\ref{1.30}) is the subject of
section III. Here we point out that these equations can be transformed into
a very useful for application form by solving for $H_n(y)$ (see Appendix A
for mathematical details): 
\begin{equation}  \label{1.31}
H_n(y)=h_n(y)+H_n,
\end{equation}
\begin{equation}  \label{1.32}
h_n(y)=\frac{\epsilon ^2}2\sum_{m=1}^{N-1}G(n,m)\frac{d\phi _m(y) }{dy},
\end{equation}
\begin{equation}  \label{1.33}
H_n=\frac{H\left( \mu ^{-n}+\mu ^{-N+n}-\mu ^n-\mu ^{N-n}\right) }{\mu
^{-N}-\mu ^N},
\end{equation}
where $G(n,m)$ are given by (\ref{b.9}), and $\mu $ is given by (\ref{b.5}).
By (\ref{b.8}), and (\ref{1.20}), (\ref{b.13}), expression (\ref{1.31})
explicitly satisfies boundary conditions (\ref{1.19}) and $H_n(\pm L)=H$.
Moreover, the $y$-independent quantities $H_n$ in (\ref{1.31}) have clear
physical meaning: Being solutions of (\ref{1.30}) with $\frac{d\phi _m}{dy}%
\equiv 0$, they describe distribution of the local magnetic field within the
I-layers in the homogeneous Meissner state. (See section III of Ref. \cite
{K00}.) [Note that in an infinite layered superconductor $H_n\equiv 0$, and $%
\sum\limits_{m=1}^{N-1}G(n,m)\ldots \rightarrow \sum\limits_{m=-\infty
}^{+\infty }G_\infty (n,m)\ldots ,$ where $G_\infty (n,m)$ are defined by (%
\ref{b.15}).] By comparing Eqs. (\ref{1.31}) with Eqs. (\ref{1.22}) (where $%
f_n=1$), using the property $H_n=H_{N-n}$ and (\ref{b.12}), we establish the
symmetry relations 
\begin{equation}  \label{1.51}
\phi _n(y)=\phi _{N-n}(y),\quad h_n(y)=h_{N-n}(y),\quad H_n(y)=H_{N-n}(y).
\end{equation}
Relations (\ref{1.51}) replace the relations $\phi _n(y)=\phi
_{n+1}(y)\equiv \phi (y)$ and $h_n(y)=h_{n+1}(y)\equiv h(y)$ of the infinite
layered superconductor.\cite{K00}

By the use of Eqs. (\ref{1.31}), the Gibbs free energy up to first order in $%
r\left( T\right) \ll 1$ can be given the form
\[
\Omega (H)=-\frac{NW}2+r(T)\left[ \frac{2H^2W}{H_s^2}\left( N-1\right)
\right. 
\]
\[
+\frac{\epsilon ^2}2\sum_{n=1}^{N-1}\sum_{m=1}^{N-1}G(n,m)\int%
\limits_{-L}^Ldy\frac{d\phi _n(y)}{dy}\frac{d\phi _m(y)}{dy} 
\]
\begin{equation}  \label{1.24.1}
\left. +2\sum_{n=1}^{N-1}\int\limits_{-L}^Ldy\sin {}^2\frac{\phi _n(y)}2%
-4H\sum_{n=1}^{N-1}\Phi _n\frac{\phi _n(L)-\phi _n(-L)}{2\pi }\right] ,
\end{equation}
\begin{equation}  \label{1.24.3}
\Phi _n=\pi \left[ 1-\frac{\mu ^{-n}+\mu ^{-N+n}-\mu ^n-\mu ^{N-n}}{\mu
^{-N}-\mu ^N}\right] ,
\end{equation}
where $H_s$ is the superheating (penetration) field defined in Eq. (\ref
{b.14}), and $\Phi _n$ is the total flux carried by a Josephson vortex
positioned in the $n$th I-layer. (The quantities $H_s$ and $\Phi _n$ are
thoroughly discussed in section IIIB.) The quadratic form of the
electromagnetic energy [the second line on the right-hand side of Eq. (\ref
{1.24.1})] can be diagonalized with the help of (\ref{b.16}), (\ref{b.17}): 
\begin{equation}  \label{1.24.2}
E_{em}=\frac 1N\sum_{k=0}^{\left[ \frac N2\right] -1}\lambda
_{2k+1}^2\sum_{n=1}^{N-1}\sum_{m=1}^{N-1}\sin \frac{\pi n\left( 2k+1\right) }%
N\sin \frac{\pi m\left( 2k+1\right) }N\int\limits_{-L}^Ldy\frac{d\phi _n(y)}{%
dy}\frac{d\phi _m(y)}{dy},
\end{equation}
where $\left[ u\right] $ is the integer part of $u$. [In deriving (\ref
{1.24.2}), we employed the symmetry relations (\ref{1.51}).] Equation (\ref
{1.24.2}) explicitly shows that physical solutions in a stack with $N$
S-layers are characterized by $\left[ \frac N2\right] $ Josephson length
scales $\lambda _{Jk}\equiv \lambda _{2k+1}$ ($k=0,1,\ldots ,\left[ \frac N2%
\right] -1$) from the set (\ref{b.16}).

\section{Major results}

\subsection{Analysis of the equations for the phase differences}

By differentiation with respect to $y$, integrodifferential equations (\ref
{1.30}) reduce to a system of $N-1$ ordinary nonlinear second-order
differential equations
\[
\frac{d^2\phi _1(y)}{dy^2}=\frac 1{\epsilon ^2}\left[ \left( 2+\epsilon
^2\right) \sin \phi _1(y)-\sin \phi _2(y)\right] , 
\]
\[
\frac{d^2\phi _n(y)}{dy^2}=\frac 1{\epsilon ^2}\left[ \left( 2+\epsilon
^2\right) \sin \phi _n(y)-\sin \phi _{n+1}(y)-\sin \phi _{n-1}(y)\right]
,\quad 2\leq n\leq N-2, 
\]
\begin{equation}  \label{1.34}
\frac{d^2\phi _{N-1}(y)}{dy^2}=\frac 1{\epsilon ^2}\left[ \left( 2+\epsilon
^2\right) \sin \phi _{N-1}(y)-\sin \phi _{N-2}(y)\right]
\end{equation}
with boundary conditions (\ref{1.20}). As shown in section II, physical
solutions to (\ref{1.34}) necessarily obey the symmetry relations (\ref{1.23}%
) and (\ref{1.51}).

A detailed analysis of the analytical structure of the coupled static SG
equations (\ref{1.34}) is the main task of this subsection. Such
differential equations were first derived within the framework of a less
rigorous, phenomenological approach in Ref. \cite{SBP93} . Particular
numerical solutions (for $H=0$) were obtained in several publications.\cite
{SBP93,KW97,Kr00,Kr01} However, any analytical investigation of these
equations has not been carried out up until now.

\subsubsection{Existence and uniqueness of the solution to the initial value
problem}

Consider Eqs. (\ref{1.34}) on the whole axis $-\infty <y<+\infty $. Two
simple properties of (\ref{1.34}) are quite obvious: If $\phi _n(y)$ ($1\leq
n\leq N-1$) is a solution, the functions $\bar \phi _n(y)$ given by 
\begin{equation}  \label{1.35}
\bar \phi _n(y)=\phi _n(y)+2\pi k\quad \text{(}k\text{ is an integer),}
\end{equation}
and 
\begin{equation}  \label{1.36}
\bar \phi _n(y)=\phi _n(y+c)\quad \text{(}c\text{ is an arbitrary constant)}
\end{equation}
are also solutions. [The latter is a result of the fact that $y$ does not
enter explicitly the right-hand side of (\ref{1.34}).] Our conclusions about
the solution to (\ref{1.34}) will be substantially based on another key
property, which we formulate as a lemma:

{\bf Lemma 1}. Consider an arbitrary interval $I=\left[ L_1,L_2\right] $ and 
$y_0\in I$. The initial value problem for Eqs. (\ref{1.34}) with arbitrary
initial conditions $\phi _n(y_0)=\alpha _n$, $\frac{d\phi _n}{dy}(y_0)=\beta
_n$ has a unique solution in the {\it whole} interval $I$. This solution has
continuous derivatives with respect to $y$ of arbitrary order and
continuously depends on the initial data. (For the proof of {\bf Lemma 1},
see Appendix B.)

It is worth noting that the existence and uniqueness of a smooth solution to
the initial value problem in the {\it whole} interval $I$ is rather
nontrivial for {\it nonlinear} differential equations: For such equations,
theorems of existence and uniqueness are usually valid only locally, in the
neighborhood of initial data.\cite{T61} In our case, global character of the
solution and its infinite differentiability are ensured by the fact that $%
\phi _n$ enter the right-hand side of Eqs. (\ref{1.34}) only as arguments of
the sine. Note that because of the arbitrariness of the interval $I$, the
solution can be uniquely continued onto the whole axis $-\infty <$$y<+\infty 
$.\cite{T61}

Differentiating (\ref{1.31}) with respect to $y$ yields 
\begin{equation}  \label{1.37}
\sin \phi _n(y)=\epsilon ^2\sum_{m=1}^{N-1}G(n,m)\frac{d^2\phi _m(y)}{dy^2}.
\end{equation}
Multiplying (\ref{1.37}) by $\frac{d\phi _n(y)}{dy}$, summing over the layer
index $n$ with the use of (\ref{b.11}) and performing integration, we arrive
at the first integral of Eqs. (\ref{1.34}): 
\begin{equation}  \label{1.38}
C\left( H\right) -\sum_{n=1}^{N-1}\cos \phi _n(y)=\frac{\epsilon ^2}2%
\sum_{n=1}^{N-1}\sum_{m=1}^{N-1}G(n,m)\frac{d\phi _n(y)}{dy}\frac{d\phi _m(y)%
}{dy},
\end{equation}
\begin{equation}  \label{1.38.1}
C\left( H\right) =\frac{2H^2}{H_s^2}\left( N-1\right) +\sum_{n=1}^{N-1}\cos
\phi _n(L),
\end{equation}
where $C\left( H\right) $ is the constant of integration. Using (\ref{1.38}%
), one can eliminate the electromagnetic-energy term from the Gibbs free
energy (\ref{1.24.1}).

The existence of the first integral of Eqs. (\ref{1.34}) was first
established in Ref. \cite{KW97}. However, the explicit form of the matrix
elements $G(n,m)$ and the actual value of $C\left( H\right) $ was not
determined by the authors of Ref. \cite{KW97}. Unfortunately, the existence
of the first integral for the {\it infinite} ($N=\infty $) layered
superconductor was not noticed in any previous publications, which partly
explains the difficulties with the minimization of the Gibbs free energy.
\cite{B73,CC90,Ca91,Ko93} [Equation (\ref{1.38}) holds for $N=\infty $ too,
taking account of the substitution $\sum\limits_{m,n=1}^{N-1}G(n,m)\ldots
\rightarrow \sum\limits_{m,n=-\infty }^{+\infty }G_\infty (n,m)\ldots $,
where $G_\infty (n,m)$ are defined by (\ref{b.15})]. In the case $N=\infty $%
, the Gibbs free energy $\Omega \left( H\right) $ should be additionally
minimized with respect to the {\it phases} $\varphi _n$, which by the use of
the first integral immediately yields the exact solution\cite{K99,K00} with $%
\phi _n(y)=\phi _{n+1}(y)\equiv \phi (y)$.

For our further consideration, we will need an auxiliary lemma:

{\bf Lemma 2}. Consider Eqs. (\ref{1.34}) on the whole axis $-\infty <$$%
y<+\infty $. Conditions 
\begin{equation}  \label{L2.1}
\phi _n(y_0)=0,\qquad \frac{d\phi _n}{dy}(y_0)=0,\qquad n=1,2,\ldots ,N-1,
\end{equation}
where $y_0\neq \pm \infty $, cannot by satisfied simultaneously for all $n$
by any nontrivial solution of (\ref{1.34}).

The proof of {\bf Lemma 2} is straightforward: Indeed, conditions (\ref{L2.1}%
) specify the initial value problem for (\ref{1.34}). This initial value
problem has the trivial solution $\phi _1=\phi _2=\ldots =\phi _{N-1}\equiv
0 $. By {\bf Lemma 1}, this solution is unique and there are no nontrivial
solutions that satisfy (\ref{L2.1}).

The importance of {\bf Lemma 2 }lies in the fact that it prohibits the
existence of any topological (vortex) solutions in any finite interval $%
\left[ L_1,L_2\right] $ in the absence of an external field ($H=0$). Indeed,
any such solution must necessarily satisfy the boundary conditions\cite
{DEGM82} $\phi _n(L_1)=0$ and $\phi _n(L_2)=0$mod$2\pi $ for all $n$, with $%
\phi _n(L_2)\neq 0$ for at least one $n=m$. For $H=0$, we also have $\frac{%
d\phi _n}{dy}(L_1)=\frac{d\phi _n}{dy}(L_2)=0$ for all $n$. However, the
former and the latter conditions are incompatible by virtue of {\bf Lemma 2}.

Another important consequence of {\bf Lemma 2} is that for any finite
interval $\left[ -L,L\right] $ the constant of integration in (\ref{1.38})
satisfies the inequality 
\begin{equation}  \label{1.38.2}
C(H)>N-1,
\end{equation}
for any $H>0$. Indeed, for the Meissner solution we have $\phi _n(y)=-\phi
_n(-y)$ and $\phi _n(0)=0$ [see (\ref{1.23})]. By setting $y=0$ in (\ref
{1.38}), applying {\bf Lemma 2} and using the fact that the quadratic form
on the right-hand side of (\ref{1.38}) is positively definite, we get (\ref
{1.38.2}). For topological solutions, inequality (\ref{1.38.2}) is quite
obvious.

\subsubsection{Localized solutions. The criterion of existence}

Alongside solutions to (\ref{1.34}) in a finite interval $\left[ -L,L\right] 
$, we will consider physical solutions to (\ref{1.34}) in the infinite ($%
-\infty <y<+\infty $) and a semiinfinite ($0\leq y<+\infty $) intervals. To
ensure the convergence of the integrals over $y$ in the expression for the
free energy [see (\ref{1.24.1})], such solutions must necessarily satisfy
the asymptotic boundary conditions\cite{DEGM82,R82} 
\begin{equation}
\phi _{n}(y)%
\mathrel{\mathop{\longrightarrow }\limits_{y\longrightarrow \pm \infty }}%
0\ \text{mod\ }2\pi ,  \label{1.40}
\end{equation}
\begin{equation}
\frac{d\phi _{n}(y)}{dy}%
\mathrel{\mathop{\longrightarrow }\limits_{y\longrightarrow \pm \infty }}%
0,  \label{1.41}
\end{equation}
for any $1\leq n\leq N-1$. (Note that {\bf Lemma 2} does not obtain for $%
y_{0}=\pm \infty $.) These solutions with a square-integrable derivative
will be called {\it localized}. Their consideration can be reduced to the
consideration of the standard initial value problem at $y=0$, with the
initial data subject to a certain existence condition.

By inserting (\ref{1.40}) and (\ref{1.41}) into (\ref{1.38}), we find the
value of the constant of integration: $C\left( 0\right) =N-1$. [Compare with
inequality (\ref{1.38.2}) for $L<\infty $, $H>0$.] Thus Eq. (\ref{1.38})
becomes 
\begin{equation}  \label{1.42}
\sum_{n=1}^{N-1}\sin {}^2\frac{\phi _n(y)}2=\frac{\epsilon ^2}4%
\sum_{n=1}^{N-1}\sum_{m=1}^{N-1}G(n,m)\frac{d\phi _n(y)}{dy}\frac{d\phi _m(y)%
}{dy}.
\end{equation}
Substituting initial values $\phi _n(0)=\alpha _n$, $\frac{d\phi _n}{dy}%
(0)=\beta _n$ into (\ref{1.42}), we obtain the desired condition on $\alpha
_n$ and $\beta _n$ that we call the {\it criterion of the existence }of
localized solutions: 
\begin{equation}  \label{1.43}
\sum_{n=1}^{N-1}\sin {}^2\frac{\alpha _n}2=\frac{\epsilon ^2}4%
\sum_{n=1}^{N-1}\sum_{m=1}^{N-1}G(n,m)\beta _n\beta _m.
\end{equation}
Indeed, {\bf Lemma 1} guarantees the existence, uniqueness and
differentiability of a solution for arbitrary $\alpha _n$ and $\beta _n$ in
the whole interval $(-\infty ,+\infty )$ [or $[0,+\infty )$]. Owing to our
special choice of the constant of integration in (\ref{1.42}), the solution
determined by $\alpha _n$ and $\beta _n$ obeying (\ref{1.43}) will
necessarily satisfy asymptotic conditions (\ref{1.40}), (\ref{1.41}).
Moreover, this solution will automatically satisfy the conditions 
\begin{equation}  \label{1.44}
\frac{d^k\phi _n(y)}{dy^k}%
\mathrel{\mathop{\rightarrow }\limits_{y\rightarrow \pm \infty }}%
0,
\end{equation}
for any $1\leq n\leq N-1$ and $2\leq k$, which can be verified by repeated
differentiation of (\ref{1.37}) and application of (\ref{1.40}), (\ref{1.41}%
).

Note that localized solutions are characterized by the absence of any
electromagnetic effects at $y\rightarrow \pm \infty $. Furthermore, owing to
(\ref{1.42}), the Josephson energy of localized solutions $E_J$ is exactly
equal to their electromagnetic energy $E_{em}$. (From a field-theoretical
point of view, this fact is a manifestation of the virial theorem.\cite{R82})

\subsubsection{Solutions with periodic derivatives. The major {\bf Theorem}}

Our conclusions about the type and the properties of physical solutions for $%
H>0$ will be based on the major {\bf Theorem}:

{\bf Theorem}. Consider Eqs. (\ref{1.34}) on the whole axis $-\infty <$$%
y<+\infty $. The initial value problem 
\begin{equation}  \label{T.1}
\phi _n(y_0)=0,\qquad \frac{d\phi _n}{dy}(y_0)=2H>0,\qquad n=1,2,\ldots
,N-1,\qquad
\end{equation}
for (\ref{1.34}) has a unique solution on the whole axis $-\infty <$$%
y<+\infty $. This solution is characterized by the properties 
\begin{equation}  \label{T.2}
\phi _n\left( y+P\right) =\phi _n\left( y\right) +2\pi ,
\end{equation}
\begin{equation}  \label{T.3}
\frac{d\phi _n}{dy}\left( y+P\right) =\phi _n\left( y\right) >0,
\end{equation}
\begin{equation}  \label{T.4}
\phi _n\left( y_0+\frac 12P\right) =\pi ,
\end{equation}
\begin{equation}  \label{T.5}
\frac{d\phi _n}{dy}\left( y_0+\frac 12P\right) =2\sqrt{H^2+H_s^2},\qquad 
\text{for all }n=1,2,\ldots ,N-1,
\end{equation}
where $H_s$ is defined via (\ref{b.14}), and the period $P$ is given by 
\begin{equation}  \label{T.6}
P(H)=\frac 2{\sqrt{H^2+H_s^2}}\int\limits_{\left( 0,0,\ldots ,0\right)
}^{\left( \frac \pi 2,\frac \pi 2,\ldots ,\frac \pi 2\right) } \frac{ds}{%
\sqrt{1-\frac 1{N-1}\frac{H_s^2}{H^2+H_s^2}\sum\limits_{n=1}^{N-1}\sin
{}^2\theta _n}},
\end{equation}
\[
ds=\epsilon H_s\sqrt{\frac 1{N-1}\sum_{n=1}^{N-1}\sum_{m=1}^{N-1}G(n,m)d%
\theta _nd\theta _m}. 
\]
The integration in (\ref{T.6}) is performed along the integral curve: 
\[
\theta _n=\frac{\phi _n(y)}2,\qquad y_0\leq y\leq y_0+\frac 12P,\qquad
n=1,2,\ldots ,N-1, 
\]
with $ds$ being the length increment. (See Appendix C for the proof.)

An immediate physical consequence of the {\bf Theorem} is the absence of
single-vortex solutions to (\ref{1.34}) in a finite interval $\left[ -L,L%
\right] $ for any $H>0$. (For $H=0$, the absence of single-vortex solutions
in a finite interval follows from {\bf Lemma 2}.) Indeed, if such solutions
existed, they would satisfy at a certain $H=H_1>0$ the boundary conditions
\cite{DEGM82}
\[
\phi _n(-L)=0,\qquad \frac{d\phi _n}{dy}(-L)=2H_1>0,\qquad n=1,2,\ldots
,N-1, 
\]
\[
\phi _l(L)=2\pi ,\quad \frac{d\phi _l}{dy}(L)=2H_1>0;\quad \phi
_n(L)=0,\quad \frac{d\phi _n}{dy}(L)=2H_1>0,\quad n\neq l. 
\]
However, these boundary conditions are mutually incompatible: The boundary
conditions at $y=-L$ specify the initial value problem for (\ref{1.34}). By
the {\bf Theorem}, this initial value problem admits a unique solution with $%
\frac{d\phi _n}{dy}>0$ on the {\it whole} axis $-\infty <$$y<+\infty $ for 
{\it all} $n$, whereas the boundary conditions at $y=L$ imply that $\frac{%
d\phi _n}{dy}$ with $n\neq l$ change the sign twice in the interval $\left[
-L,L\right] $.

Analogously, one can prove the absence for $H>0$ and $L<\infty $ of any
other incoherent vortex solutions, as well as such vortex configurations as
the ''triangular lattice''\cite{V89,Ca91,BC91} and the ''row mode''.\cite
{Kr00} In contrast, the properties (\ref{T.2})-(\ref{T.6}) are inherent to
the vortex-plane solutions.\cite{K99,K00}

Note that $P(H)$ $%
\mathrel{\mathop{\rightarrow }\limits_{H\rightarrow 0}}%
\infty $, in agreement with {\bf Lemma 2}. On the other hand, for $H\gg H_s$%
, the path of integration in (\ref{T.6}) is a straight line: $\theta _n=%
\frac \pi P\left( y-y_0\right) $, $y_0\leq y\leq y_0+\frac 12P$, $%
n=1,2,\ldots ,N-1$, which yields $P(H)$ $%
\mathrel{\mathop{\rightarrow }\limits_{H\gg H_s}}%
$ $\frac \pi H$.

For a single thin-layer junction ($N=2$), equations (\ref{1.34}) reduce to a
single SG equation for $\phi _1\equiv \phi $, with $\lambda
_{J0}=H_s^{-1}\equiv \frac \epsilon {\sqrt{2+\epsilon ^2}}$. In agreement
with Refs. \cite{K70,OS67} , the period (\ref{T.6}) becomes 
\begin{equation}  \label{T.7}
P(H)=\frac 2{\sqrt{H^2+H_s^2}}K\left( \frac{H_s^2}{H^2+H_s^2}\right) ,
\end{equation}
where $K\left( k^2\right) $ is the complete elliptic integral of the first
kind.\cite{AS65} Relations (\ref{T.2})-(\ref{T.6}), applied to $\phi $,
determine an {\it exact} vortex solution in terms of the Jakobi elliptic
functions am$(u)$ and dn$(u)=\frac d{du}$am$(u)$.\cite{AS65}

The true vortex-plane solutions appear in a double-junction stack ($N=3$).
Owing to the symmetry relations (\ref{1.51}), equations (\ref{1.34}) again
reduce to a single SG equation for $\phi _1=\phi _2\equiv \phi $, with $%
\lambda _{J0}=H_s^{-1}\equiv \frac \epsilon {\sqrt{1+\epsilon ^2}}$. The
period $P$ is again given by (\ref{T.7}) (with redefined $H_s$), and the 
{\it exact} vortex-plane solution is again expressed via the functions am$%
(u) $ and dn$(u)$. These two exactly-solvable examples ($N=2$ and $N=3$) not
only demonstrate the power and generality of the {\bf Theorem}, but also
show that the vortex-plane solutions are a natural generalization of the
well-known vortex solutions in a single junction.\cite{K70,OS67,BP82} (See
section IV for a more detailed discussion of the cases $N=2,3$.)

\subsection{Meissner solutions. The superheating (penetration) fields $%
H_{sL} $ and $H_s\equiv H_{s\infty }$}

In a finite interval $\left[ -L,L\right] $, the Meissner solution is
characterized by the relations 
\begin{equation}  \label{1.51.1}
\phi _n(y)=-\phi _n(-y),\qquad 1\leq n\leq N-1,
\end{equation}
resulting from the general symmetry (\ref{1.23}), and obeys the conditions (%
\ref{1.51}). Thus, the Meissner boundary value problem is completely
specified by the boundary conditions 
\begin{equation}  \label{1.46.0}
\frac{d\phi _n}{dy}\left( -L\right) =2H>0,\qquad \phi _n\left( 0\right)
=0,\qquad 1\leq n\leq N-1.\quad
\end{equation}
Up to a certain field $H=H_{sL}$, the boundary value problem (\ref{1.46.0})
has a unique solution with $\frac{d\phi _n}{dy}>0$ in the whole interval $%
\left[ -L,L\right] $ and $-\pi \leq \phi _n\left( -L\right) <0$ for all $n$.
The field $H_{sL}$ is determined from the conditions $\phi _n\left( \pm
L\right) =\pm \pi $ for all $n$. [As follows from (\ref{1.38.1}), (\ref
{1.38.2}), $H_{sL}>H_s$.] Indeed, for $H\ll H_s$ the Meissner solution can
be obtained in a closed analytical form as a linear combination of $\left[ 
\frac N2\right] $ exponentials with $\lambda _{Jk}$ [see (\ref{1.24.2})]:
According to (\ref{1.38.1}), (\ref{1.38.2}), this case allows of
linearization. The existence of a unique Meissner solution for higher values
of $H$ follows from continuous dependence of the solution on the initial
data $\phi _n\left( -L\right) =\alpha _n$ and $\frac{d\phi _n}{dy}\left(
-L\right) =2H\equiv \beta _n$. (See {\bf Lemma 1}.) The Meissner solution
ceases to exist at $H=H_{sL}$: For this field, both $\frac{d\phi _n}{dy}(y)$
and $h_n(y)$ are local maxima at $y=\pm L$. [See (\ref{1.32}), (\ref{1.37}).]

Physically, the conditions $\phi _n\left( \pm L\right) =\pm \pi $ correspond
to the vanishing of the surface barrier induced by Josephson currents [$%
j_{n,n-1}\left( \pm L\right) =0$] and the formation of a vortex plane at the
side boundaries of the stack.\cite{K99} (Thus, at $H=H_{sL}$, the total flux
due to the field penetration through the $y=\pm L$ interfaces is exactly
equal to the flux carried by a single vortex plane.) For $H>H_{sL}$, only
topological vortex-plane solutions are possible. However, the vortex-plane
solutions can exist and be favorable energetically already at fields $%
H<H_{sL}$. [See the next subsection.] For these reasons, $H_{sL}$ should be
identified both with the {\it superheating} field of the Meissner state and
the {\it penetration} field for the vortex planes. According to the {\bf %
Theorem}, at $H=H_{sL}$, $\frac{d\phi _n}{dy}(0)=2\sqrt{H_{sL}^2-H_s^2}$ for
all $n$, and hence $H_n(0)=\sqrt{H_{sL}^2-H_s^2}+\left( H_{sL}-\sqrt{%
H_{sL}^2-H_s^2}\right) \left[ G(n,1)+G(n,N-1)\right] $. The field $H_{sL}$
itself is determined by the implicit equation 
\begin{equation}  \label{1.46.01}
2L=P\left( \sqrt{H_{sL}^2-H_s^2}\right) ,
\end{equation}
where $P$ is given by (\ref{T.6}), with the path of integration being
determined by the solution to the boundary value problem $\phi _n\left(
-L\right) =-\pi $, $\phi _n\left( 0\right) =0$. Note that $H_{sL}\rightarrow
H_s$ for $L\gg H_s^{-1}$, which will be explained below.

The consideration of the Meissner effect becomes simpler, when $L\gg \lambda
_{J\max }\equiv \lambda _{J0}$. In this case, the interval $\left[ -L,L%
\right] $ can be transformed into $[0,+\infty )$ by changing the variable $%
y\rightarrow y-L$ and proceeding to the limit $L\rightarrow \infty $. In the
semiinfinite interval $[0,+\infty )$, the Meissner solution necessarily
satisfies the conditions $\phi _n(y)%
\mathrel{\mathop{\rightarrow }\limits_{y\rightarrow +\infty }}%
0$, $\frac{d\phi _n(y)}{dy}%
\mathrel{\mathop{\rightarrow }\limits_{y\rightarrow +\infty }}%
0$, and we can use the criterion (\ref{1.43}) that now takes the form 
\begin{equation}  \label{1.53}
\frac 1{N-1}\sum_{n=1}^{N-1}\sin {}^2\frac{\phi _n(0)}2=\frac{H^2 }{H_s^2},
\end{equation}
where, by (\ref{b.14}) and (\ref{b.16}), (\ref{b.17}), 
\[
H_s=\left[ 1-\frac{\left( 2\sqrt{1+\frac{\epsilon ^2}4}-\epsilon \right)
\left( 1-\mu ^{N-1}\right) }{\epsilon \left( N-1\right) \left( 1+\mu
^{N-1}\right) }\right] ^{-\frac 12} 
\]
\begin{equation}  \label{1.54}
\equiv \sqrt{\frac{\left( N-1\right) N}2}\left[ \sum_{k=0}^{\left[ \frac N2%
\right] -1}\lambda _{Jk}^2\cot {}^2\frac{\pi \left( 2k+1\right) }{2N}\right]
^{-\frac 12}.
\end{equation}
Physical interpretation of the quantity $H_s$ is straightforward. The
maximum value of the left-hand side of (\ref{1.53}) is achieved when all $%
\phi _n(0)=-\pi $ and is equal to unity, which corresponds to $H=H_s$ on the
right-hand side. Hence $H_s$ should be identified with the superheating
(penetration) field $H_{s\infty }$ ($H_s\equiv H_{s\infty }$). Note that $%
H_s $ for $N<\infty $ is always higher than the corresponding field\cite
{K99,K00} $\bar H_s=1$ of the infinite ($N=\infty $) layered superconductor.
For $N\rightarrow \infty $, $H_s\rightarrow 1$.

According to (\ref{1.53}), equations (\ref{1.34}) can be linearized for $%
H\ll H_s$. In this limit, the explicit Meissner solution in the interval $%
[0,+\infty )$ is 
\begin{equation}  \label{1.54.1}
\phi _n(y)=-\frac{4H}N\sum_{k=0}^{\left[ \frac N2\right] -1}\lambda
_{Jk}\cot \frac{\left( 2k+1\right) \pi }{2N}\sin {}\frac{\left( 2k+1\right)
n\pi }N\exp \left[ -\frac y{\lambda _{Jk}}\right] ,
\end{equation}
\begin{equation}  \label{1.54.2}
H_n(y)=\frac{2H}N\sum_{k=0}^{\left[ \frac N2\right] -1}\lambda _{Jk}^2\cot 
\frac{\left( 2k+1\right) \pi }{2N}\sin {}\frac{\left( 2k+1\right) n\pi }N%
\exp \left[ -\frac y{\lambda _{Jk}}\right] +H_n,
\end{equation}
where $H_n$ is given by (\ref{1.33}). The distribution of the currents $J_n$
and $j_{n,n-1}$ can be easily obtained from (\ref{1.25}), (\ref{1.26}) using
(\ref{1.54.1}), (\ref{1.54.2}).

It is instructive to investigate a transition to the infinite
layered-superconductor limit $N-1\gg 2\left[ \epsilon ^{-1}\right] $ ($%
\epsilon <1$) in Eqs. (\ref{1.54.1}), (\ref{1.54.2}). Thus, for I-layers
whose layer index $n$ satisfies the condition 
\begin{equation}  \label{1.54.3}
\left[ \epsilon ^{-1}\right] \ll n\ll N-1-\left[ \epsilon ^{-1}\right] ,
\end{equation}
we have $H_n=H\left( \mu ^n+\mu ^{N-n}\right) \ll H$. Under the condition (%
\ref{1.54.3}), the main contribution to the sums over $k$ in (\ref{1.54.1}),
(\ref{1.54.2}) comes from $k\ll \left[ \frac N2\right] -1$. For such $k$ and 
$N\gg 1$, $\lambda _{Jk}\approx \lambda _{J0}\approx \bar \lambda _J=1$.
Therefore, for $n$ in the interval (\ref{1.54.3}) we get
\[
\phi _n(y)\approx -\frac{8H}\pi \exp \left( -y\right) \sum_{k=0}^{+\infty } 
\frac{\left( -1\right) ^k}{2k+1}=-2H\exp \left( -y\right) , 
\]
\[
H_n(y)\approx H\exp \left( -y\right) , 
\]
in complete agreement with the results of Refs. \cite{K99,K00} .

Two final remarks would be in order here. The fact that the Meissner
solution in a stack with $N$ S-layers is characterized by $\left[ \frac N2%
\right] $ Josephson lengths $\lambda _{Jk}$ was not noticed in any previous
publications. Moreover, in contrast to continuum type-II superconductors,
\cite{dG} the local field $H_n(y)$ at $H=H_s$ cannot be represented as a sum
of a purely Meissner field and a vortex field, because the principle of
superposition does not hold for the nonlinear Eqs. (\ref{1.34}).
Unfortunately, the latter point was not realized in Ref. \cite{BF92}
concerned with infinite layered superconductors.

\subsection{Vortex-plane solutions. The lower critical field $H_{c1}$}

As explained in the Introduction, by a vortex plane we understand a coherent
chain of $N-1$ Josephson vortices (one vortex per each I-layer) positioned
in a plane parallel to the coordinate axes $x$, $z$: see Fig. 2. (In this
plane, all $\frac{d\phi _n}{dy}$ and $h_n$ are local maxima.) Such solutions
in an interval $\left[ -L,L\right] $ obey the general relations (\ref{1.23})
with the {\it same} constant $0$mod$2\pi \neq 0$ for {\it all} $n$ and the
symmetry conditions (\ref{1.51}): hence the same set of $\left[ \frac N2%
\right] $ Josephson lengths $\lambda _{Jk}$ as in the Meissner case. The 
{\bf Theorem} envisages the existence of vortex-plane solutions for any $H>0$%
. Using the {\bf Theorem} and {\bf Lemma 1}, we can easily establish all
basic properties of the vortex-plane solutions. They can be summarized as
follows.

The boundary value problem for the vortex-plane solutions is specified by
the boundary conditions 
\begin{equation}  \label{1.46}
\frac{d\phi _n}{dy}\left( -L\right) =2H>0,\qquad \phi _n\left( 0\right) =\pi
N_v,\qquad 1\leq n\leq N-1,\qquad N_v=1,2,\ldots ,\quad
\end{equation}
where $N_v$ is the number of vortex planes. As in the case of the Meissner
boundary value problem (\ref{1.46.0}), the boundary value problem (\ref{1.46}%
) for a fixed $N_v$ and any $L>0$ has a unique solution, with $\frac{d\phi
_n }{dy}>0$ in the whole interval $\left[ -L,L\right] $ and $-\pi \leq \phi
_n\left( -L\right) \leq 0$ for all $n$, in a certain field range 
\begin{equation}  \label{1.46.1}
\sqrt{H_{N_v-1}^2-H_s^2}\leq H\leq H_{N_v},
\end{equation}
where $H_0\equiv H_{sL}$. The lower bound of the existence of the $N_v$%
-vortex-plane solution is determined by the boundary value problem 
\begin{equation}  \label{1.46.2}
\phi _n\left( -L\right) =0,\qquad \phi _n\left( 0\right) =\pi N_v,\qquad
1\leq n\leq N-1,
\end{equation}
whereas the upper bound is determined by the boundary value problem 
\begin{equation}  \label{1.46.3}
\phi _n\left( -L\right) =-\pi ,\qquad \phi _n\left( 0\right) =\pi N_v,\qquad
1\leq n\leq N-1.
\end{equation}
The field $H_{N_v}$ is determined by the implicit equation 
\begin{equation}  \label{1.46.4}
2L=\left( N_v+1\right) P\left( \sqrt{H_{N_v}^2-H_s^2}\right) ,
\end{equation}
where $P$ is given by (\ref{T.6}), with the path of integration being
determined by the solution to the boundary value problem (\ref{1.46.3}).
Note that for $N_v=0$ (the Meissner state) Eq. (\ref{1.46.4}) coincides with
Eq. (\ref{1.46.01}), as it should. Physically, $P$ determines the period of
the distribution of $H_n(y)$ and $j_{n,n-1}(y)$.

As follows from (\ref{1.46.1}), the range of the existence of the state with 
$N_v$ vortex planes overlaps with that of the state with $N_v-1$ vortex
planes. Thus, the state with $N_v=1$ overlaps with the Meissner state. For $%
L\gg H_s^{-1}$, the overlapping can occur for several neighboring states.
The overlapping is negligibly small only for $L\ll H_s^{-1}$ (with arbitrary 
$N_v$) and for $N_v\gg 1$ (with arbitrary $L$). (For ordinary Josephson
vortices in a single junction, the overlapping is well known.\cite{K70,OS67}%
) However, the actual number of vortex planes, $N_v$, for given $H$ should
be determined from the requirement that the free energy be an absolute
minimum. Note that for $L\rightarrow \infty $, $\sqrt{H_0^2-H_s^2}\equiv 
\sqrt{H_{sL}^2-H_s^2}\rightarrow 0$ and the state with $N_v=1$ appears
already in zero external field. This case is discussed in more detail below.

For $L\gg H_s^{-1}$, we can proceed to the limit $L\rightarrow \infty $ and
consider the state with $N_v=1$ in the interval $(-\infty ,+\infty )$. With
the asymptotic boundary conditions 
\begin{equation}  \label{1.56}
\phi _n(y)%
\mathrel{\mathop{\rightarrow }\limits_{y\rightarrow -\infty }}%
0,\quad \phi _n(y)%
\mathrel{\mathop{\rightarrow }\limits_{y\rightarrow +\infty }}%
2\pi ,\qquad \frac{d\phi _n(y)}{dy}%
\mathrel{\mathop{\rightarrow }\limits_{y\rightarrow \pm \infty }}%
0
\end{equation}
for all $n$, this state satisfies the criterion (\ref{1.43}) and can be
obtained as the solution to an initial value problem at $y=0$. Indeed, it
can be constructed from the Meissner solution in the interval $[0,+\infty )$
at $H=H_s$: Using the property (\ref{1.35}), by means of the transformation $%
\phi _n\rightarrow \phi _n+2\pi $ we obtain in the interval $[0,+\infty )$ a
solution that satisfies the initial conditions 
\begin{equation}  \label{1.57.1}
\alpha _n\equiv \phi _n(0)=\pi ,\quad \beta _n\equiv \frac{d\phi _n}{dy}%
(0)=2H_s,\quad 1\leq n\leq N-1,
\end{equation}
and the conditions (\ref{1.56}) for $y\rightarrow +\infty $. By {\bf Lemma 1}%
, this solutions can be uniquely continued into the whole interval $(-\infty
,+\infty )$. The solution thus obtained satisfies the conditions (\ref{1.56}%
) for $y\rightarrow \pm \infty $ and hence represents the desired
singe-vortex-plane solution. By the construction, $h_n(0)=H_s\left[
1-G(n,1)-G(n,N-1)\right] $. The total flux carried by the vortex plane is 
\begin{equation}  \label{1.60}
\Phi =\sum_{n=1}^{N-1}\Phi _n=\pi \left( N-1\right) \left[ 1- \frac{2\sqrt{1+%
\frac{\epsilon ^2}4}-\epsilon }{\epsilon \left( N-1\right) } \frac{1-\mu
^{N-1}}{1+\mu ^N}\right] ,
\end{equation}
where $\Phi _n$ is given by (\ref{1.24.3}). Note that in contrast to
Josephson junctions with thick electrodes\cite{BP82} and infinite layered
superconductors,\cite{K99,K00} the flux carried by a Josephson vortex in a
finite thin-layer S/I structure {\it is not quantized} and is always smaller
than the flux quantum $\Phi _0=\pi $. For $N-1\gg 2\left[ \epsilon ^{-1}%
\right] $, $\Phi \rightarrow \pi \left( N-1\right) $, as it should.

To determine the thermodynamic {\it lower critical }field $H_{c1}$ at which
the vortex-plane solutions for $L\gg H_s^{-1}$ become energetically
favorable, we must calculate the difference between the Gibbs free energy in
the presence of a single vortex plane, $\Omega _v(H)$, and the Gibbs free
energy of the homogeneous Meissner state, $\Omega _M(H)$ [the sum of the
phase-independent terms in (\ref{1.24.1})]: 
\[
\Omega _v(H)-\Omega _M(H) 
\]
\begin{equation}  \label{1.62}
=r(T)\left[ \epsilon
^2\sum_{n=1}^{N-1}\sum_{m=1}^{N-1}G(n,m)\int\limits_{-\infty }^{+\infty }dy 
\frac{d\phi _n(y)}{dy}\frac{d\phi _m(y)}{dy}-4\Phi H\right] ,
\end{equation}
where the total flux $\Phi $ is given by (\ref{1.60}). The first term on the
right-hand side of (\ref{1.62}) should be interpreted as the self-energy of
the vortex plane: $E_v=2E_{em}=2E_J$. From (\ref{1.62}), we get: 
\begin{equation}  \label{1.64}
H_{c1}=\frac{E_v}{4r(T)\Phi }.
\end{equation}
For thermodynamically stable solutions, we must necessarily have $H_{c1}<H_s$%
. It is straightforward to verify that this condition is met by the
vortex-plane solutions. Using (\ref{1.37}), (\ref{1.56}), the initial values 
$\beta _n=2H_s$ and integrating by parts, we convert $E_v$ into the form 
\begin{equation}  \label{1.65}
E_v=2r(T)\left[ \sum_{n=1}^{N-1}\int\limits_{+\infty }^0dy\phi _n(y)\sin
\phi _n(y)-2H_s\Phi \right] .
\end{equation}
The first term on the right-hand side of (\ref{1.65}) is positive, because
in the region $0\leq y<+\infty $ all $\phi _n$ satisfy the relation $\pi
\leq \phi _n<2\pi $. By the use of (\ref{1.37}) and (\ref{b.14}), we obtain
the following strict inequalities:
\[
2H_s\Phi <\sum_{n=1}^{N-1}\int\limits_{+\infty }^0dy\phi _n(y)\sin \phi
_n(y)<4H_s\Phi , 
\]
\[
0<E_v<4r(T)H_s\Phi . 
\]
Hence, 
\[
0<H_{c1}<H_s, 
\]
as anticipated. Note that in all special cases admitting exact analytical
solutions ($N=\infty $,\cite{K99,K00} and $N=2,3$, see section IV), $H_{c1}=%
\frac 2\pi H_s$.

\subsection{Single-vortex solutions for $H=0$, $L=\infty $, and other
localized incoherent vortex solutions}

As shown in section IIIA, the only topological (vortex) solutions admitted
by Eqs. (\ref{1.34}) for $H>0$ are the vortex-plane solutions. For $H=0$,
equations (\ref{1.34}) do not possess any topological solutions in a finite
interval $\left[ -L,L\right] $. Therefore, incoherent vortex solutions to (%
\ref{1.34}) can exist only for $H=0$, in the infinite interval $(-\infty
,+\infty )$, and must necessarily meet the requirements imposed on localized
solutions. Below we consider in detail the most important type of such
solutions, namely single-vortex solutions.

A single Josephson vortex positioned in the $l$th I-layer at $y=0$ obeys the
symmetry relations 
\begin{equation}  \label{1.68}
\phi _l(y)=2\pi -\phi _l(-y);\quad \phi _n(y)=-\phi _n(-y),\quad n\neq l,
\end{equation}
[see (\ref{1.23})] and the asymptotic boundary conditions 
\begin{equation}  \label{1.69}
\phi _l(y)%
\mathrel{\mathop{\rightarrow }\limits_{y\rightarrow -\infty }}%
0,\quad \phi _l(y)%
\mathrel{\mathop{\rightarrow }\limits_{y\rightarrow +\infty }}%
2\pi ;\qquad \phi _n(y)%
\mathrel{\mathop{\rightarrow }\limits_{y\rightarrow \pm \infty }}%
0,\quad n\neq l,
\end{equation}
\begin{equation}  \label{1.71}
\frac{d\phi _n(y)}{dy}%
\mathrel{\mathop{\rightarrow }\limits_{y\rightarrow \pm \infty }}%
0,\quad \text{for all }1\leq n\leq N-1.
\end{equation}
Moreover, $\frac{dh_n(y)}{dy}>0$ in the region $-\infty <y<0$, and $\frac{%
dh_n(y)}{dy}<0$ in the region $0<y<+\infty $. Hence, $\phi _n$ must satisfy
the relations 
\begin{equation}  \label{1.72}
0<\phi _n(y)<\pi ,\text{ }y\in \left( -\infty ,0\right) ;\quad -\pi <\phi
_n(y)<0,\text{ }y\in \left( 0,+\infty \right) ,\quad \text{for }n\neq l,
\end{equation}
and the initial conditions 
\begin{equation}  \label{1.73}
\alpha _l\equiv \phi _l(0)=\pi ;\quad \alpha _n\equiv \phi _n(0)=0,\quad
n\neq l,
\end{equation}
\begin{equation}  \label{1.74}
\beta _l\equiv \frac{d\phi _l(0)}{dy}>0;\quad \beta _n\equiv \frac{d\phi
_n(0)}{dy}<0,\quad n\neq l.
\end{equation}
A necessary condition of the existence of such a solution is provided by the
general criterion (\ref{1.43}) and has the form 
\begin{equation}  \label{1.75}
\frac{\epsilon ^2}4\sum_{n=1}^{N-1}\sum_{m=1}^{N-1}G(n,m)\beta _n\beta _m=1.
\end{equation}

The flux through the $n$th I-layer due to the vortex in the $l$th I-layer
can be found using (\ref{1.32}) and (\ref{1.69}): 
\begin{equation}  \label{1.76}
\Phi _{nl}=\int\limits_{-\infty }^{+\infty }dyh_n(y)=\frac{\pi \epsilon }{2%
\sqrt{1+\frac{\epsilon ^2}4}}\left[ \mu ^{\left| n-l\right| }- \frac{\mu
^n\left( \mu ^{l-N}-\mu ^{N-l}\right) +\mu ^{N-n}\left( \mu ^{-l}-\mu
^l\right) }{\mu ^{-N}-\mu ^N}\right] .
\end{equation}
The total flux carried by the vortex in the $l$th layer is $\Phi
_l=\sum_{n=1}^{N-1}\Phi _{nl}$, where $\Phi _l$ is given by (\ref{1.24.3})
with $n=l$. For $N<\infty $, the total flux $\Phi _l$ is not quantized and
is less than the flux quantum $\Phi _0=\pi $. (See the previous subsection.)

In contrast to the Meissner and the vortex-plane solutions, single-vortex
solutions, in general, do not obey the symmetry (\ref{1.51}), inherent to
the original integrodifferential equations (\ref{1.18}), (\ref{1.30})
minimizing the Gibbs free-energy. Therefore, they are characterized by the
full set (\ref{b.16}) of $N-1$ length scales $\lambda _i$. (The only
exclusion is a stack with an odd number of junctions $N-1$ and $l=\frac N2$.)

Although the energy of single-vortex solutions is lower than the self-energy
of the vortex-plane solutions at $H=0$, the former solutions cannot be used,
e.g., for estimates of $H_{c1}$ because of their absolute instability at $%
H>0 $. Unfortunately, this crucial issue was not understood in any previous
publications. Thus, the main distinctive feature of single-vortex solutions
that precludes their existence at $H>0$, i.e., $\frac{d\phi _n(0)}{dy}<0$
for $n\neq l$, was overlooked in Refs. \cite{B73,CC90,Ko93} concerned with
infinite layered superconductors. (The approach of Refs. \cite{B73,CC90,Ko93}
was criticized from different points of view in Refs. \cite{F98,K99,K00} .)
The negative sign of $\frac{d\phi _n(0)}{dy}$ for $n\neq l$ was noticed in
Ref. \cite{Ca91} . However, the effect of this property on the stability of
single-vortex solutions at $H>0$ was not realized therein.

The feature (\ref{1.74}) is clearly reproduced by the numerical studies\cite
{SBP93,Kr00,Kr01} of finite stacks. Unfortunately, in the absence of any
analytical investigations of Eqs. (\ref{1.34}), the authors of Refs. \cite
{SBP93,Kr00,Kr01} failed to establish the actual domain of validity of their
numerical results: $H=0$, $L=\infty $. The condition (\ref{1.75}) was not
found either. In this regard, it would be appropriate to point out an
intrinsic limitation of the numerical calculations that does not allow one
to consider them as a proof of the existence of single-vortex solutions even
in the case $H=0.$ By necessity, these calculations are performed in a {\it %
finite} interval $\left[ L_1,L_2\right] $. However, as shown in section
IIIA, equations (\ref{1.34}) do not admit {\it any} topological solutions
(in a strict mathematical sense\cite{DEGM82,R82}) in any finite interval at $%
H=0$. It should be also noted that the single-vortex solution cannot be
represented by a linear combination of the terms $4\arctan \exp \left[ \frac %
y{\lambda _i}\right] $,\cite{Kr00,Kr01} because the principle of
superposition does not apply to the nonlinear Eqs. (\ref{1.34}).

The consideration of other localized incoherent vortex solutions to (\ref
{1.34}) (i.e., solutions with $2\leq k<N-1$ vortices in the plane $y=0$, and
vortex-antivortex configurations) can be done along the same lines. At $H>0$%
, all these solutions are unstable too.

\section{The two exactly-solvable cases}

Below, we present an exact solution for the cases $N=2,3$. Valid for any $%
H\geq 0$ and $L>0$, this solution provides a clear illustration of the
general results of sections IIIA-IIIC. The single-junction case ($N=2$)
illuminates some typical specific features of the thin-S-layer limit and
establishes a profound physical and mathematical relationship between the
ordinary Josephson vortices and the vortex-plane solutions. In the
double-junction case ($N=3$), our exact solution describes the true vortex
planes, which is of considerable experimental interest: For a
double-junction stack, we already have direct observations of
Josephson-vortex configurations that can be unambiguously identified with
the vortex-plane solutions.\cite{N96}

\subsection{A single thin-layer junction ($N=2$)}

In this simplest case, a single phase difference $\phi _1(y)\equiv \phi (y)$
satisfies the usual static SG equation 
\begin{equation}  \label{4.2}
\frac{d^2\phi (y)}{dy^2}=\frac 1{\lambda _{J0}^2}\sin \phi (y),
\end{equation}
with the Josephson length\cite{K70} 
\begin{equation}  \label{4.3}
\lambda _{J0}=\frac \epsilon {\sqrt{2+\epsilon ^2}}.
\end{equation}
Note that $\lambda _{J0}$, given by (\ref{4.3}), for $\epsilon \ll 1$ is
much smaller than the Josephson length of a single junction with thick
electrodes, which in our dimensionless units is $\lambda _J=\sqrt{\frac p{%
2\lambda }}$.\cite{K70} From (\ref{1.33}) and (\ref{1.54}), we get the local
field in the homogeneous Meissner state 
\begin{equation}  \label{4.4}
H_1=\frac{2H}{2+\epsilon ^2},
\end{equation}
and the superheating (penetration) field 
\begin{equation}  \label{4.5}
H_s=\lambda _{J0}^{-1}=\frac{\sqrt{2+\epsilon ^2}}\epsilon ,
\end{equation}
respectively. For $\epsilon \ll 1$, the superheating (penetration) field $%
H_s $, given by (\ref{4.5}), is much higher than the corresponding field\cite
{K70,BP82} of a single junction with thick electrodes $H_s=\lambda _J$.

Using the {\bf Theorem}, we immediately find an exact solution to the
boundary value problems (\ref{1.46.0}) and (\ref{1.46}): 
\begin{equation}  \label{n1}
\phi (y)=\pi \left( N_v-1\right) +2\text{am}\left( \frac y{k\lambda _{J0}}%
+K\left( k^2\right) ,k\right) ,
\end{equation}
\begin{equation}  \label{n2}
\text{dn}\left( \frac L{k\lambda _{J0}},k\right) =\frac{\sqrt{1-k^2}}{%
k\lambda _{J0}H},\qquad N_v=2m,\quad m=0,1,\ldots ;
\end{equation}
\begin{equation}  \label{n3}
\phi (y)=\pi N_v+2\text{am}\left( \frac y{k\lambda _{J0}},k\right) ,
\end{equation}
\begin{equation}  \label{n4}
\text{dn}\left( \frac L{k\lambda _{J0}},k\right) =k\lambda _{J0}H,\qquad
N_v=2m+1,\quad m=0,1,\ldots ,
\end{equation}
where am$\left( u\right) $ and dn$\left( u\right) =\frac d{du}$am$\left(
u\right) $ are the Jakobi elliptic functions, and $K\left( k^2\right) $ is
the elliptic integral of the first kind.\cite{AS65} The range of the
existence of the solution with $N_v=0,1,\ldots $ Josephson vortices is given
by (\ref{1.46.1}), where $H_{N_v}$ is determined by the implicit equation 
\begin{equation}  \label{n5}
\frac L{\lambda _{J0}}=\frac{\left( N_v+1\right) }{H_{N_v}\lambda _{J0}}%
K\left( \frac 1{H_{N_v}^2\lambda _{J0}^2}\right) .
\end{equation}

Note that although the Meissner and the vortex solutions to (\ref{4.2}) were
studied a long time ago,\cite{K70,OS67} a closed-form analytical solution of
the type (\ref{n1})-(\ref{n5}), valid for any $0<L$ in the whole field range 
$0\leq H\ll H_{c2}$, has not been found up until now. (Apparently for this
reason, there exists an erroneous belief\cite{BCG92,Kr01} that Josephson
vortices ''do not form'' in single junctions with small $L$.) Using
asymptotics of am$\left( u\right) $, dn$\left( u\right) $ and $K\left(
k^2\right) $,\cite{AS65} we can obtain from (\ref{n1})-(\ref{n5}) all
physically interesting limiting cases.

\subsubsection{The Meissner solution for $[0\leq y<+\infty )$}

For the fields $0\leq H\leq H_s=\frac{\sqrt{2+\epsilon ^2}}\epsilon $, the
Meissner solution in the semiinfinite interval $[0,+\infty )$ can be
obtained from (\ref{n1}), (\ref{n2}) with $N_v=0$ by changing the variable $%
y\rightarrow y-L$ and proceeding to the limit $L\rightarrow \infty $. Using
the formulas of section II, up to first order in $r(T)\ll 1$, we have 
\begin{equation}  \label{4.6}
\phi (y)=-4\arctan \frac{H\exp \left[ -\frac y{\lambda _{J0}}\right] }{H_s+%
\sqrt{H_s^2-H^2}},
\end{equation}
\begin{equation}  \label{4.7}
H(y)\equiv H_1(y)=h(y)+H_1,
\end{equation}
\begin{equation}  \label{4.8}
h(y)\equiv h_1(y)=\frac{2\lambda _{J0}H\left[ H_s+\sqrt{H_s^2-H^2}\right]
\exp \left[ -\frac y{\lambda _{J0}}\right] }{\left[ H_s+\sqrt{H_s^2-H^2}%
\right] ^2+H^2\exp \left[ -\frac{2y}{\lambda _{J0}}\right] },
\end{equation}
\[
j(y)\equiv j_{1,0}(y)=-4H\left[ H_s+\sqrt{H_s^2-H^2}\right] 
\]
\begin{equation}  \label{4.9}
\times \frac{\left[ \left[ H_s+\sqrt{H_s^2-H^2}\right] ^2-H^2\exp \left[ -%
\frac{2y}{\lambda _{J0}}\right] \right] \exp \left[ -\frac y{\lambda _{J0}}%
\right] }{\left[ \left[ H_s+\sqrt{H_s^2-H^2}\right] ^2+H^2\exp \left[ - 
\frac{2y}{\lambda _{J0}}\right] \right] ^2},
\end{equation}
\begin{equation}  \label{4.10}
J(y)\equiv J_0(y)=J_1(y)=\frac 1{4\pi }\left[ H-H_1-h(y)\right] ,
\end{equation}
\begin{equation}  \label{4.11}
f(y)\equiv f_0(y)=f_1(y)=1-\frac{r(T)}2\left[ \lambda _{J0}^{-2}h(y)+\frac 2{%
\epsilon ^2}\left[ h(y)+H_1-H\right] ^2\right] .
\end{equation}

\subsubsection{The single-vortex solution for $(-\infty <y<+\infty )$}

The single-vortex solution in the infinite interval $(-\infty ,+\infty )$
can be obtained from (\ref{n3}), (\ref{n4}) with $N_v=1$ by proceeding to
the limit $L\rightarrow \infty $. It has the form 
\begin{equation}  \label{4.16}
\phi (y)=4\arctan \exp \left[ \frac y{\lambda _{J0}}\right] ,
\end{equation}
\begin{equation}  \label{4.17}
h(y)=\lambda _{J0}\cosh {}^{-1}\left[ \frac y{\lambda _{J0}}\right] ,
\end{equation}
\begin{equation}  \label{4.18}
j(y)=-2\cosh {}^{-2}\left[ \frac y{\lambda _{J0}}\right] \sinh {}\left[ 
\frac y{\lambda _{J0}}\right] .
\end{equation}
The quantities $H(y)$, $J(y)$ and $f(y)$ are given by (\ref{4.7}), (\ref
{4.10}) and (\ref{4.11}), respectively, with $h(y)$ taken from (\ref{4.17}).

By inserting (\ref{4.16}) into (\ref{1.62}) with $H=0$, we obtain the vortex
self-energy: 
\begin{equation}  \label{4.18.1}
E_v=8r(T)\frac \epsilon {\sqrt{2+\epsilon ^2}}.
\end{equation}
The vortex flux, according to (\ref{1.60}), is 
\[
\Phi =\pi \frac{\epsilon ^2}{2+\epsilon ^2}, 
\]
and the lower critical field, by (\ref{1.64}), is 
\begin{equation}  \label{4.19}
H_{c1}=\frac 2\pi H_s=\frac 2\pi \frac{\sqrt{2+\epsilon ^2}}\epsilon .
\end{equation}
Thus, for $\epsilon \ll 1$, the vortex flux $\Phi \ll \Phi _0=\pi $, and the
lower critical field (\ref{4.19}) is much larger than the corresponding
field of a single junction with thick electrodes $H_{c1}=\frac 2\pi \sqrt{%
\frac p{2\lambda }}$, in agreement with Ref. \cite{KW97}.

\subsubsection{The vortex solution for small Josephson screening}

When the screening by Josephson currents is negligibly small, i.e., {\it (i)}
for $L\ll \lambda _{J0}$ and arbitrary $H$, or {\it (ii)} for $H_s\ll H\ll
H_{c2}$ and arbitrary $L$, equations (\ref{n1})-(\ref{n5}) become
\[
\phi (y)=\pi N_v+2Hy 
\]
\begin{equation}  \label{n6}
-\frac{\left( -1\right) ^{N_v}}{4\lambda _{J0}^2H^2}\left[ \sin \left(
2Hy\right) -2Hy\cos \left( HW\right) \right] ,
\end{equation}
where $N_v=\left[ \frac{HW}\pi \right] $. Equations of the type (\ref{n6})
were first obtained for the infinite layered superconductor by means of a
perturbation theory.\cite{K99,K00} As can be seen from (\ref{n6}), the
overlapping of states with different $N_v$ now vanishes, and the period of
the vortex structure is given by $P=\frac \pi H$, in full agreement with the
general results of sections IIIA and IIIC.

\subsection{A double-junction stack ($N=3$)}

Owing to the symmetry (\ref{1.51}), $\phi _1(y)=\phi _2(y)\equiv \phi (y)$,
and a double-junction stack is described by the single SG equation (\ref{4.2}%
) with the single Josephson length 
\begin{equation}  \label{4.21}
\lambda _{J0}=\frac \epsilon {\sqrt{1+\epsilon ^2}}.
\end{equation}
An exact solution to the boundary value problems (\ref{1.46.0}) and (\ref
{1.46}) is again given by Eqs. (\ref{n1})-(\ref{n5}), where now $N_v$ should
be interpreted as the number of vortex planes. Thus, practically all the
formulas of the previous subsection (with corresponding redefinition of the
physical quantities) hold for the double-junction stack too. Note that
\[
\phi _1(y)=\phi _2(y)\equiv \phi (y),\quad H_1(y)=H_2(y)\equiv H(y),\quad
h_1(y)=h_2(y)\equiv h(y), 
\]
\[
j_{1,0}(y)=j_{2,1}(y)\equiv j(y),\quad J_0(y)=J_2(y)\equiv J(y),\quad
f_0(y)=f_2(y)\equiv f(y), 
\]
\[
J_1(y)=0,\quad f_1(y)=1-\frac{r(T)}{\lambda _{J0}^2}h(y),\quad H_1=H_2=\frac %
H{1+\epsilon ^2}. 
\]
According to (\ref{1.54}), the superheating (penetration ) field is 
\[
H_s=\lambda _{J0}^{-1}=\frac{\sqrt{1+\epsilon ^2}}\epsilon , 
\]
which is smaller than the corresponding single-junction value (\ref{4.5}),
in agreement with Ref. \cite{GGU96}.

The vortex-plane self-energy for the interval $(-\infty ,+\infty )$ is 
\begin{equation}  \label{4.28}
E_v=16r(T)\lambda _{J0}=16r(T)\frac \epsilon {\sqrt{1+\epsilon ^2}},
\end{equation}
and the flux is 
\[
\Phi =2\pi \frac{\epsilon ^2}{1+\epsilon ^2}, 
\]
which immediately leads to the lower critical field: 
\[
H_{c1}=\frac 2\pi H_s=\frac 2\pi \frac{\sqrt{1+\epsilon ^2}}\epsilon . 
\]
As can be seen by comparing (\ref{4.28}) with the single-junction expression
(\ref{4.18.1}), the energy per vortex in the double-junction stack is
higher. The vortex-plane solutions are presented schematically in Fig. 2.

\section{Discussion}

Within the framework of standard methods of the theory of differential
equations, we have obtained a complete mathematical description of the
Meissner effect and the vortex structure in periodic thin-layer $(N-1)$%
-junction stacks ($2\leq N<\infty $). The results of our analytical analysis
of the coupled static SG equations (\ref{1.34}), summarized in {\bf Lemmas 1}%
,{\bf 2} and the {\bf Theorem}, should provide a basis for any further
analytical or numerical study of these equations, not necessarily restricted
to the field of superconductivity.

By proving the absence of single-vortex solutions to (\ref{1.34}) for $H>0$
and specifying the actual domain of their existence ($H=0$, $L=\infty $), we
have clarified a wide-spread misunderstanding:\cite{B73,CC90,Ca91,Ko93} The
previous estimates of $H_{c1}$ turn out to be irrelevant because of absolute
thermodynamic instability of single-vortex configurations. In full agreement
with our study of infinite ($N=\infty $) layered superconductors,\cite
{K99,K00} we have shown that the only true topological (vortex)
configurations that survive at $H>0$ are the vortex planes. Being a natural
generalization of ordinary Josephson vortices in a single junction ($N=2$),
the vortex-plane solutions for $3\leq N$ inherit such properties of the
former as periodicity along the layers [Eq. (\ref{T.2})] and the overlapping
of states with different topological numbers $N_v$ [Eq. (\ref{1.46.1})]. A
unified mathematical approach of this paper, valid for $2\leq N<\infty $,
allowed us to derive an exact expression [Eq. (\ref{1.54})] for the
superheating (penetration) field $H_s$ as an explicit function of $N$.
Within the framework of the same unified approach, we have obtained a new
exact solution to the single SG equation, Eqs. (\ref{n1})-(\ref{n5}), valid
for any $H\geq 0$ and $L>0$. This solution refutes the assertions\cite
{BCG92,Kr01} that Josephson vortices ''do not form'' in single junctions
with small $L$.

Our closed-form analytical expression for the Meissner field, Eq. (\ref
{1.54.2}), clearly illustrates an important feature of the Meissner effect
in $(N-1)$-junction stacks, not noticed in previous theoretical
publications, namely the existence of $\left[ \frac N2\right] $ different
Josephson lengths $\lambda _{Jk}$ ($k=0,1,\ldots ,\left[ \frac N2\right] -1$%
). This result may prove to be useful in view of the current experimental
efforts\cite{MKHLX98,PCXWC00} to verify the interlayer tunneling model of
high-$T_c$ superconductivity\cite{WHA88} by measuring the $c$-axis
penetration depth. [The penetration of the parallel magnetic field with a
distribution of length scales has been recently observed\cite{KMSW99} in the
organic layered superconductor $\kappa $-(BEDT-TTF)$_2$Cu(NCS)$_2$.]

Finally, the vortex-plane solutions of our paper provide a natural
explanation for the observed coherent Josephson-vortex configurations in
artificial double-junction stacks\cite{N96} and weakly-coupled multilayers
\cite{Yu98} in the presence of a static parallel external field $H>0$. The
observation of isolated interlayer vortices in the high-$T_c$ superconductor
Tl$_2$Ba$_2$CuO$_{6+\delta }$\cite{MKHLX98} and in $\kappa $-(BEDT-TTF)$_2$%
Cu(NCS)$_2$\cite{KMSW99} can be explained by the fact that in these
experiments the external field was set equal to zero ($H=0$). In this
situation, isolated Josephson vortices can be pinned by structural defects,
because their self-energy is lower and the spatial extension along the $c$%
-axis is smaller than those of the vortex planes at $H=0$. We hope that our
results will stimulate further experimental investigations.

\appendix

\section{The solution of the finite difference equation for $H_n(y)$}

Equations (\ref{1.30}) can be regarded as a nonhomogeneous finite difference
equation for $H_n(y)$ with respect to the layer index $n$, subject to
boundary conditions (\ref{1.19}). According to general theory of such
equations,\cite{G59} its solution can be represented in the form 
\begin{equation}  \label{b.1}
H_n(y)=h_n(y)+H_n,
\end{equation}
where 
\begin{equation}  \label{b.2}
h_n(y)=\frac{\epsilon ^2}2\sum_{m=1}^{N-1}G(n,m)\frac{d\phi _m(y) }{dy}
\end{equation}
is the particular solution of (\ref{1.30}) satisfying the boundary
conditions 
\begin{equation}  \label{b.3}
h_0(y)=h_N(y)=0,
\end{equation}
and 
\begin{equation}  \label{b.4}
H_n=\frac{H\left( \mu ^{-n}+\mu ^{-N+n}-\mu ^n-\mu ^{N-n}\right) }{\mu
^{-N}-\mu ^N},
\end{equation}
\begin{equation}  \label{b.5}
\mu =1+\frac{\epsilon ^2}2-\epsilon \sqrt{1+\frac{\epsilon ^2}4}, \text{ (}%
0<\mu <1\text{),}
\end{equation}
is the solution of the homogeneous form of (\ref{1.30}) (with the zero
right-hand side) meeting the boundary conditions 
\begin{equation}  \label{b.6}
H_0=H_N=H.
\end{equation}

The quantities $G(n,m)$ in (\ref{b.2}) are elements of a $\left( N-1\right)
\times \left( N-1\right) $ matrix ${\bf G(}1\leq n,m\leq N-1)$. They obey
the nonhomogeneous finite difference equation 
\begin{equation}  \label{b.7}
\left( 2+\epsilon ^2\right) G(n,m)-G(n+1,m)-G(n-1,m)=\delta _{n,m}
\end{equation}
($\delta _{n,m}$ is the Kronecker index) with the boundary conditions 
\begin{equation}  \label{b.8}
G(0,m)=G(N,m)=0.
\end{equation}
The explicit form of $G(n,m)$ is 
\begin{equation}  \label{b.9}
G(n,m)=\frac 1{2\epsilon \sqrt{1+\frac{\epsilon ^2}4}}\left[ \mu ^{\left|
n-m\right| }-\frac{\mu ^n\left( \mu ^{m-N}-\mu ^{N-m}\right) +\mu
^{N-n}\left( \mu ^{-m}-\mu ^m\right) }{\mu ^{-N}-\mu ^N}\right] .
\end{equation}
The following properties of $G(n,m)$ can be easily verified using (\ref{b.7}%
) and (\ref{b.9}): 
\begin{equation}  \label{b.11}
G(n,m)=G(m,n),
\end{equation}
\begin{equation}  \label{b.12}
G(n,N-m)=G(N-n,m),
\end{equation}
\begin{equation}  \label{b.10}
G(n,m)>0\text{ for any }1\leq n,m\leq N-1,
\end{equation}
\[
\sum_{m=1}^{N-1}G(n,m)=\frac 1{\epsilon ^2}\left[ 1-G(n,1)-G(n,N-1)\right] 
\]
\begin{equation}  \label{b.13}
=\frac 1{\epsilon ^2}\left[ 1-\frac{\mu ^{-n}+\mu ^{-N+n}-\mu ^n-\mu ^{N-n}}{%
\mu ^{-N}-\mu ^N}\right] ,\quad 1\leq n\leq N-1,
\end{equation}
\begin{equation}  \label{b.14}
\sum_{n=1}^{N-1}\sum_{m=1}^{N-1}G(n,m)=\frac 1{\epsilon ^2}\left[ N-1-\frac{2%
\sqrt{1+\frac{\epsilon ^2}4}-\epsilon }\epsilon \frac{1-\mu ^{N-1}}{1+\mu ^N}%
\right] \equiv \frac{N-1}{\epsilon ^2H_s^2}.
\end{equation}
According to (\ref{b.7}), the inverse of the matrix ${\bf G(}n,m)$ is a
Jacobian matrix\cite{GK50} ${\bf G^{-1}(}n,m)$ with elements $%
G^{-1}(n,n)=2+\epsilon ^2$ ($n=1,2,\ldots ,N-1$), $%
G^{-1}(n+1,n)=G^{-1}(n,n+1)=-1$ ($n=1,2,\ldots ,N-2$), and $G^{-1}(n,m)=0$
for $\left| n-m\right| >1$. Hence all the eigenvalues $e_j$ of ${\bf G(}n,m)$
are positive and given by 
\begin{equation}  \label{b.16}
e_j=\frac{\lambda _j^2}{\epsilon ^2},\qquad \lambda _j=\frac \epsilon {\sqrt{%
2+\epsilon ^2-2\cos \frac{\pi j}N}},\qquad i=1,2,\ldots ,N-1,
\end{equation}
with the corresponding eigenvectors 
\begin{equation}  \label{b.17}
u_{ij}=\sqrt{\frac 2N}\sin \frac{\pi ij}N,\qquad i,j=1,2,\ldots ,N-1.
\end{equation}
The quantities $\lambda _j$ determine the characteristic length scales of
the stack with $N$ S-layers.

Note that for an infinite layered superconductor, the analog of the matrix $%
{\bf G(}n,m)$ is ${\bf G}_\infty (n,m)$, determined by the matrix elements 
\begin{equation}  \label{b.15}
G_\infty (n,m)=\frac{\mu ^{\left| n-m\right| }}{2\epsilon \sqrt{1+\frac{%
\epsilon ^2}4}},\quad -\infty <n,m<+\infty ,
\end{equation}
that satisfy the summation rule 
\[
\sum_{m=1}^{N-1}G_\infty (n,m)=\frac 1{\epsilon ^2}. 
\]

\section{Proof of {\bf Lemma 1}}

By introducing new functions 
\[
\psi _1(y)=\phi _1(y),\psi _2(y)=\phi _2(y),\ldots ,\psi _{N-1}(y)=\phi
_{N-1}(y), 
\]
\begin{equation}  \label{a.1}
\psi _N(y)=\frac{d\phi _1(y)}{dy},\psi _{N+1}(y)=\frac{d\phi _2(y) }{dy}%
,\ldots ,\psi _{2N-2}(y)=\frac{d\phi _{N-1}(y)}{dy},
\end{equation}
we convert (\ref{1.34}) into an equivalent normal system of $2N-2$
first-order equations 
\begin{equation}  \label{a.2}
\frac{d\psi _i(y)}{dy}=F_i\left( \psi _1,\psi _2,\ldots ,\psi _{2N-2}\right)
,\quad 1\leq i\leq 2N-2,
\end{equation}
\[
F_i\left( \psi _1,\psi _2,\ldots ,\psi _{2N-2}\right) \equiv \psi
_{i+N-1},\quad 1\leq i\leq N-1, 
\]
\[
F_N\left( \psi _1,\psi _2,\ldots ,\psi _{2N-2}\right) \equiv \frac 1{%
\epsilon ^2}\left[ \left( 2+\epsilon ^2\right) \sin \psi _1-\sin \psi _2%
\right] , 
\]
\[
F_i\left( \psi _1,\psi _2,\ldots ,\psi _{2N-2}\right) \equiv \frac 1{%
\epsilon ^2}\left[ \left( 2+\epsilon ^2\right) \sin \psi _i-\sin \psi
_{i-1}-\sin \psi _{i+1}\right] ,\quad N+1\leq i\leq N-3, 
\]
\[
F_{2N-2}\left( \psi _1,\psi _2,\ldots ,\psi _{2N-2}\right) \equiv \frac 1{%
\epsilon ^2}\left[ \left( 2+\epsilon ^2\right) \sin \psi _{N-1}-\sin \psi
_{N-2}\right] , 
\]
subject to initial conditions
\[
\psi _i(y_0)=\alpha _i,\quad 1\leq i\leq N-1, 
\]
\begin{equation}  \label{a.3}
\psi _i(y_0)=\beta _{i-N+1},\quad N\leq i\leq 2N-2.
\end{equation}

To prove the statement of {\bf Lemma 1}, it is sufficient to observe that
all $F_i\left( \psi _1,\psi _2,\ldots ,\psi _{2N-2}\right) $ are continuous
functions of their arguments for $y\in (-\infty ,+\infty )$ and $\psi _k\in
(-\infty ,+\infty )$ ($1\leq k\leq 2N-2$). Moreover, their partial
derivatives with respect to $\psi _k$ satisfy the relation 
\begin{equation}  \label{a.4}
\left| \frac{\partial F_i\left( \psi _1,\psi _2,\ldots ,\psi _{2N-2}\right) 
}{\partial \psi _k}\right| \leq \frac{4+\epsilon ^2}{\epsilon ^2}
\end{equation}
for $y\in (-\infty ,+\infty )$ and $\psi _k\in (-\infty ,+\infty )$ ($1\leq
i,k\leq 2N-2$). Thus, the Lipschitz conditions with respect to $\psi _k$ are
met for $y\in (-\infty ,+\infty )$ and $\psi _k\in (-\infty ,+\infty )$ ($%
1\leq k\leq 2N-2$), which immediately guarantees\cite{T61} the existence and
uniqueness of a solution to (\ref{a.2}), satisfying arbitrary initial
conditions (\ref{a.3}), in an arbitrary interval $I=\left[ L_1,L_2\right] $
such that $y_0\in I$. Continuous dependence of the solution on initial data
is a result of continuous dependence of $F_i\left( \psi _1,\psi _2,\ldots
,\psi _{2N-2}\right) $ on their arguments and of the condition (\ref{a.4}).
Infinite differentiability of the solution automatically follows from
infinite differentiability of $F_i\left( \psi _1,\psi _2,\ldots ,\psi
_{2N-2}\right) $ with respect to their arguments.

\section{Proof of the Theorem}

Owing to the property (\ref{1.36}), without loss of generality, we can
consider the initial value problem (\ref{T.1}) for $y=y_0\equiv 0$. To prove
the Theorem, we have to show that there exists a solution to (\ref{1.34}) in 
$(-\infty ,+\infty )$ with the properties (\ref{T.2})-(\ref{T.6}) that at $%
y=y_0\equiv 0$ satisfies (\ref{T.1}). By {\bf Lemma1}, this solution will
represent the sought unique solution to the initial value problem (\ref{T.1}%
).

Consider an arbitrary finite interval $I=\left[ -\frac P2,\frac P2\right] $.
We start with the Meissner boundary value problem in the interval $I$: 
\begin{equation}  \label{c.1}
\frac{d\phi _n}{dy}\left( \pm \frac P2\right) =2\tilde H>0,\qquad \phi
_n(0)=0,\qquad 1\leq n\leq N-1.
\end{equation}
Up to a certain $\tilde H=\tilde H_s$, the problem (\ref{c.1}) has a unique
solution with the properties 
\begin{equation}  \label{c.2}
\phi _n(y)=-\phi _n(-y),\qquad y\in I,\qquad 1\leq n\leq N-1,
\end{equation}
\begin{equation}  \label{c.3}
\frac{d\phi _n}{dy}\left( y\right) >0,\qquad y\in I,\qquad 1\leq n\leq N-1,
\end{equation}
\begin{equation}  \label{c.4}
-\pi \leq \phi _n\left( -\frac P2\right) <0,\qquad 0<\phi _n\left( \frac P2%
\right) \leq \pi ,\qquad 1\leq n\leq N-1,
\end{equation}
where $\tilde H_s$ is determined by the conditions 
\begin{equation}  \label{c.5}
\phi _n\left( \pm \frac P2\right) =\pm \pi ,\qquad 1\leq n\leq N-1,
\end{equation}
and satisfies the inequality $\tilde H_s>H_s$, with $H_s$ defined via (\ref
{b.14}). (See section IIIB.)

Now we will prove that {\it all} $\frac{d\phi _n}{dy}\left( 0\right) $,
where $\phi _n$ are the solution to (\ref{c.1}), for $\tilde H\rightarrow 
\tilde H_s$ tend to the {\it same} limiting value 
\begin{equation}  \label{c.6}
\frac{d\phi _n}{dy}\left( 0\right) =2\sqrt{\tilde H_s^2-H_s^2},\qquad 1\leq
n\leq N-1.
\end{equation}
Indeed, relations (\ref{c.6}) satisfy (\ref{1.38}) at $y=0$ with (\ref
{1.38.1}), where the substitution $H\rightarrow \tilde H_s$, $L\rightarrow 
\frac P2$ has been made. However, it is still necessary to show that the
limiting value (\ref{c.6}) is uniquely determined by the solution to (\ref
{c.1}). To this end, we consider the relation 
\begin{equation}  \label{c.7}
4\left( N-1\right) \frac{\tilde H}{H_s^2}-\epsilon
^2\sum_{n=1}^{N-1}\sum_{m=1}^{N-1}G(n,m)\frac{d\beta _n}{d\tilde H}\beta
_m=\sum_{n=1}^{N-1}\sin \phi _n\left( \frac P2\right) \frac{d\phi _n\left( 
\frac P2\right) }{d\tilde H}\geq 0,
\end{equation}
where $\beta _n\equiv \frac{d\phi _n}{dy}\left( 0\right) $. [Relation (\ref
{c.6}) is obtained from (\ref{1.38}), (\ref{1.38.1}) by the substitution $%
H\rightarrow \tilde H$, $L\rightarrow \frac P2$ and differentiation with
respect to $\tilde H$.] Notice that 
\begin{equation}  \label{c.8}
\beta _n<2\tilde H,\qquad 1\leq n\leq N-1,
\end{equation}
in the whole range $0<\tilde H\leq \tilde H_s$, because all $\frac{d\phi _n}{%
dy}$ have minima at $y=0$. [See (\ref{1.37}).] With increasing $\tilde H$
from zero to $\tilde H_s$, all $\phi _n\left( \frac P2\right) $ increase
monotonously from zero to $\pi $. Thus, the difference on the left-hand side
of (\ref{c.6}) first increases (for $0<\tilde H\ll H_s$) and then decreases
(for $H_s\leq \tilde H\leq \tilde H_s$). It means that the dependence $\beta
_n(\tilde H)$ changes from a linear one for $0<\tilde H\ll H_s$ [i.e., $%
\beta _n(\tilde H)=\bar b_n\tilde H$, $0<\bar b_n<2$, $1\leq n\leq N-1$] to
a nonlinear one for $H_s\leq \tilde H\leq \tilde H_s$, with 
\begin{equation}  \label{c.9}
\frac{d\beta _n}{d\tilde H}>2,\qquad 1\leq n\leq N-1.
\end{equation}
At $\tilde H=\tilde H_s$ the right-hand side of (\ref{c.6}) is equal to
zero, and we have 
\begin{equation}  \label{c.10}
\frac{d\beta _n}{d\tilde H_s}\beta _m=b_nb_m\tilde H_s,
\end{equation}
where $b_n$ satisfy the conditions 
\begin{equation}  \label{c.11}
\epsilon ^2\sum_{n=1}^{N-1}\sum_{m=1}^{N-1}G(n,m)b_nb_m=\frac{4\left(
N-1\right) }{H_s^2},
\end{equation}
\begin{equation}  \label{c.12}
0<b_{\min }\leq b_n\leq b_{\max },\quad \quad b_{\min }=%
\mathrel{\mathop{\min }\limits_{n}}%
\left\{ b_n\right\} \leq 2,\quad \quad b_{\max }=%
\mathrel{\mathop{\max }\limits_{n}}%
\left\{ b_n\right\} \geq 2.
\end{equation}
The integration of (\ref{c.10}) for $n=m$ under the condition $\beta
_n\left( H_s\right) =0$ yields 
\begin{equation}  \label{c.13}
\beta _n\left( \tilde H_s\right) =b_n\sqrt{\tilde H_s^2-H_s^2}.
\end{equation}
With the help of (\ref{c.8}) and (\ref{c.9}), taken at $\tilde H=\tilde H_s$%
, we establish the following inequalities for $b_{\min }$ and $b_{\max }:$%
\begin{equation}  \label{c.14}
2\sqrt{1-\frac{H_s^2}{\tilde H_s^2}}<b_{\min }\leq 2,\qquad 2\leq b_{\max }<%
\frac 2{\sqrt{1-\frac{H_s^2}{\tilde H_s^2}}}.
\end{equation}
Inequalities (\ref{c.14}) must hold for any $\tilde H_s>H_s$. Proceeding to
the limit $\frac{H_s^2}{\tilde H_s^2}\rightarrow 0$, we get $b_{\min
}=b_{\max }=2$, hence the relations (\ref{c.6}).

The Meissner solution at $\tilde H=\tilde H_s$ can be uniquely continued
from the interval $I$ into the whole interval $(-\infty ,+\infty )$.\cite
{T61} The solution $\phi _n$ ($1\leq n\leq N-1$) thus obtained possesses all
the required properties (\ref{T.1})-(\ref{T.6}), where $y_0\equiv 0$, and $%
H\equiv \sqrt{\tilde H_s^2-H_s^2}$. To show this, it is sufficient to prove
the property (\ref{T.2}). [The properties (\ref{T.1}), (\ref{T.4}) and (\ref
{T.5}) are obvious. The property (\ref{T.3}) results from the
differentiation of (\ref{T.2}). The property (\ref{T.6}) is obtained from (%
\ref{1.38}), (\ref{1.38.1}) by integration.]

Consider a set of functions 
\begin{equation}  \label{c.15}
\bar \phi _n(y)=\phi _n\left( y+P\right) -2\pi ,\qquad 1\leq n\leq N-1,
\end{equation}
where $\phi _n(y)$ ($1\leq n\leq N-1$) are the continuation of the Meissner
solution at $\tilde H=\tilde H_s$ into the interval $(-\infty ,+\infty \dot )
$. Owing to the properties (\ref{1.35}), (\ref{1.36}), the functions (\ref
{c.15}) also represent a solution to (\ref{1.34}) in the interval $(-\infty
,+\infty \dot )$. At $y=-\frac P2$, the solution $\bar \phi _n$ ($1\leq
n\leq N-1$) satisfies the same initial conditions as the solution $\phi _n$ (%
$1\leq n\leq N-1$). By {\bf Lemma 1}, the latter means that $\bar \phi
_n\equiv \phi _n$ ($1\leq n\leq N-1$) in the interval $(-\infty ,+\infty 
\dot )$, i.e., 
\begin{equation}  \label{c.16}
\phi _n\left( y+P\right) -2\pi =\phi _n(y),\qquad 1\leq n\leq N-1,
\end{equation}
which accomplishes the proof of the {\bf Theorem}. Note that in this proof
we have not employed the symmetry relations (\ref{1.51}) resulting from the
boundary conditions (\ref{1.19}).

\begin{center}
\newpage\ {\bf FIGURE CAPTIONS}
\end{center}

\begin{enumerate}
\item  Fig. 1. The geometry of the problem (schematically). Here $N=4$, $%
a\ll p$ ($a$ is the S-layer thickness), $2L=W$, $H>0$.

\item  Fig. 2. The vortex-plane solutions in a double-junction stack
(schematically): a) the state with a single vortex plane ($N_v=1$); b) the
state with two vortex planes ($N_v=2$). The black dots conventionally denote
the positions of Josephson vortices, and the arrows show the distribution of
intralayer and Josephson currents.
\end{enumerate}


\begin{references}

\bibitem{K99}  S. V. Kuplevakhsky, Phys. Rev. B {\bf 60}, 7496 (1999).

\bibitem{K00}  S. V. Kuplevakhsky, Phys. Rev. B {\bf 63}, 054508 (2001).

\bibitem{SBP93}  S. Sakai, P. Bodin, and N. F. Pedersen, J. Appl. Phys. {\bf %
73}, 2411 (1993).

\bibitem{LD}  W. E. Lawrence and S. Doniach, in {\it Proceedings of the
Twelfth Conference on Low Temperature Physics, Kyoto, 1970, }edited by E.
Kanda (Keigaku, Tokyo, 1970), p. 361.

\bibitem{KW97}  V. M. Krasnov and D. Winkler, Phys. Rev. B {\bf 56}, 9106
(1997).

\bibitem{Kr00}  V. M. Krasnov, Physica C {\bf 332}, 308 (2000).

\bibitem{Kr01}  V. M. Krasnov, Phys. Rev. B {\bf 63}, 064519 (2001).

\bibitem{T61}  E. A. Coddington and N Levinson, {\it Theory of Ordinary
Differential Equations} (McGraw Hill, New York, 1955); F. G. Tricomi, {\it %
Differential Equations} (Blackie and Son, 1961).

\bibitem{Th90}  S. Theodorakis, Phys. Rev. B {\bf 42}, 10172 (1990).

\bibitem{SAUK96}  S. N. Song, P. R. Auvil, M. Ulmer, and J. B. Kettterson,
Phys. Rev. B {\bf 53}, R6018 (1996).

\bibitem{GGU96}  E. Goldobin, A. Golubov, and A. V. Ustinov, Czech. J. Phys. 
{\bf 46}, Suppl. S2, 663 (1996).

\bibitem{N96}  I. P. Nevirkovets, T. Doderer, A. Laub, M. G. Blamire, and J.
E. Evetts, J. Appl. Phys. {\bf 80}, 2321 (1996).

\bibitem{Yu98}  S. M. Yusuf, E. E. Fullerton, R. M. Osgood II, and G. P.
Felcher, J. Appl. Phys. {\bf 83}, 6801 (1998); S. M. Yusuf, R. M. Osgood
III, J. S. Jiang, C. H. Sowers, S. D. Bader, E. E. Fullerton, and G. P.
Felcher, J. Magn. Magn. Mater. {\bf 198}-{\bf 199}, 564 (1999).

\bibitem{V89}  A. F. Volkov, Phys. Lett. A {\bf 138}, 213 (1989).

\bibitem{Ca91}  J. P. Carton, J. Phys. I (France) {\bf 1}, 113 (1991).

\bibitem{BC91}  L. N. Bulaevskii and J. R. Clem, Phys. Rev. B {\bf 44},
10234 (1991).

\bibitem{B73}  L. N. Bulaevskii, Zh. Eksp. Teor. Fiz. {\bf 64}, 2241 (1973)
[Sov. Phys. JETP {\bf 37}, 1133 (1973)].

\bibitem{CC90}  J. R. Clem and M. W. Coffey, Phys. Rev. B {\bf 42}, 6209
(1990); J. R. Clem, M. W. Coffey, and Z. Hao, {\it ibid.} {\bf 44}, 2732
(1991).

\bibitem{Ko93}  A. E. Koshelev, Phys. Rev. B {\bf 48}, 1180 (1993).

\bibitem{F98}  B. Farid, J. Phys. Condens. Matter {\bf 10}, L589 (1998).

\bibitem{r1}  A full derivation of Eqs. (\ref{1.16})-(\ref{1.22}) requires
minimization with respect to $f_n$ and the vector potential of a microscopic
free-energy functional:\cite{K99} See section III in Ref. \cite{K00} for
technical details.

\bibitem{DEGM82}  R. K. Dodd, J. C.Eilbeck, J. D. Gibbon, H. C. Morris, {\it %
Solitons and Nonlinear Wave Equations} (Academic Press, London, 1982).

\bibitem{R82}  R. Rajaraman, {\it Solitons and Instantons} (North-Holland,
Amsterdam, 1982).

\bibitem{K70}  I. O. Kulik, Zh. Eksp. Teor. Fiz. {\bf 51}, 1952 (1966) [Sov.
Phys. JETP {\bf 24}, 1307 (1967)]; I. O. Kulik and I. K. Yanson, {\it The
Josephson Effect in Superconducting Tunneling Structures} (Israel Program
for Scientific Translation, Jerusalem, 1972).

\bibitem{OS67}  C. S. Owen and D. J. Scalapino, Phys. Rev. {\bf 164}, 538
(1967).

\bibitem{AS65}  M. Abramowitz and I. A. Stegun, {\it Handbook of
Mathematical Functions} (Dover, New York, 1965).

\bibitem{BP82}  A. Barone and G. Paterno, {\it Physics and Applications of
the Josephson Effect} (Wiley, New York, 1982).

\bibitem{dG}  P. G. de Gennes, {\it Superconductivity of Metals and Alloys}
(Benjamin, New York, 1966).

\bibitem{BF92}  A. Buzdin and D. Feinberg, Physics Letters A {\bf 165}, 281
(1992).

\bibitem{BCG92}  L. N. Bulaevskii, J. R. Clem, and I. I. Glazman, Phys. Rev.
B {\bf 46}, 350 (1992).

\bibitem{MKHLX98}  K. A. Moler, J. R. Kirtley, D. G. Hinks, T. W. Li, and M.
Xu, Science {\bf 279}, 1193 (1998).

\bibitem{PCXWC00}  C. Panagopoulos, J. R. Cooper, T. Xiang, Y. S. Wang, and
C. W. Chu, Phys. Rev. B {\bf 61}, R3808 (2000).

\bibitem{WHA88}  J. M. Wheatley, T. Hsu, and P. W. Anderson, Nature (London) 
{\bf 333}, 121 (1988); P. W. Anderson, Science {\bf 256}, 1526 (1992); {\it %
ibid. }{\bf 256}, 1526 (1992); {\it ibid. }{\bf 268}, 1154 (1995); {\it %
ibid. }{\bf 279}, 1196 (1998); S. Chakravarty, A. Sudb\o , P. W. Anderson,
and S. Strong, {\it ibid. }{\bf 261}, 337 (1993).

\bibitem{KMSW99}  J. R. Kirtley, K. A. Moler, J. A. Schueter, and J. M.
Williams, J. Phys.: Condens. Matter {\bf 11}, 2007 (1999).

\bibitem{G59}  A. O. Gelfond, {\it The Calculus of Finite Difference}
(GIFML, Moscow, 1959) (in Russian).

\bibitem{GK50}  F. R. Gantmacher and M. G. Krejn, {\it Oscillating Matrices
and Kernels, and Small Oscillations of Mechanical Systems} (GITTL, Moscow,
1950) (in Russian).
\end{references}
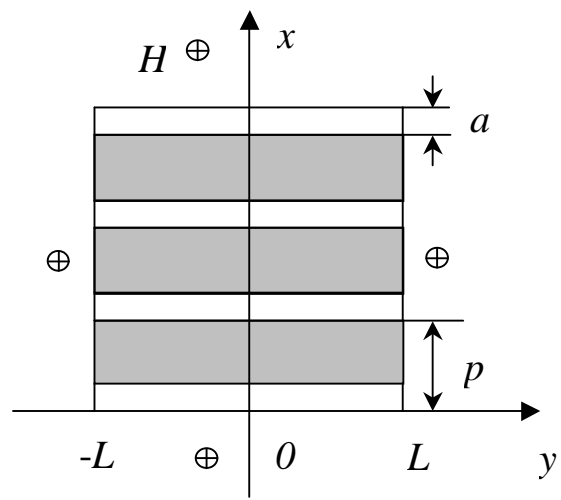
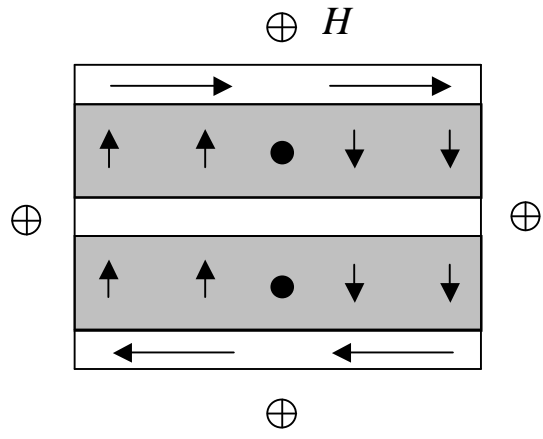
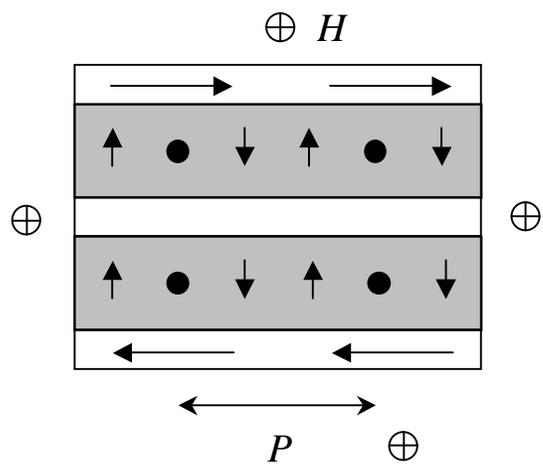
\end{document}